\pdfoutput=1

\RequirePackage{fix-cm}

\documentclass[twocolumn]{svjour3}

\smartqed

\usepackage{graphicx}
\usepackage{amsmath,amssymb}
\usepackage{txfonts}
\usepackage{float}
\usepackage[numbers]{natbib}
\usepackage[unicode]{hyperref}

\journalname{Meccanica}

\begin{document}

\title{Investigating the planar circular restricted three-body problem with strong gravitational field}

\author{Euaggelos E. Zotos}

\institute{Department of Physics, School of Science, \\
Aristotle University of Thessaloniki, \\
GR-541 24, Thessaloniki, Greece \\
Corresponding author's email: {evzotos@physics.auth.gr}}

\date{Received: 10 February 2016 / Accepted: 29 September 2016 / Published online: 8 October 2016}

\titlerunning{Investigating the planar circular restricted three-body problem with strong gravitational field}

\authorrunning{Euaggelos E. Zotos}

\maketitle

\begin{abstract}

The case of the planar circular restricted three-body problem where one of the two primaries has a stronger gravitational field with respect to the classical Newtonian field is investigated. We consider the case where two primaries have the same mass, so as the the only difference between them to be the strength of the gravitational field which is controlled by the power $p$ of the potential. A thorough numerical analysis takes place in several types of two dimensional planes in which we classify initial conditions of orbits into three main categories: (i) bounded, (ii) escaping and (iii) collision. Our results reveal that the power of the gravitational potential has a huge impact on the nature of orbits. Interpreting the collision motion as leaking in the phase space we related our results to both chaotic scattering and the theory of leaking Hamiltonian systems. We successfully located the escape as well as the collision basins and we managed to correlate them with the corresponding escape and collision time of the orbits. We hope our contribution to be useful for a further understanding of the escape and collision properties of motion in this interesting version of the restricted three-body problem.


\end{abstract}

\section{Introduction}
\label{intro}

Escaping particles from dynamical systems is a subject to which has been devoted many studies over the years. Especially the issue of escapes in Hamiltonian systems is directly related to the problem of chaotic scattering which has been an active field of research over the last decades and it still remains open (e.g., \cite{BTS96,BST98,BGOB88,BOG89,BGO90,CPR75,C90,CK92,E88,JS88,ML02,NH01,OT93,PH86,SASL06,SSL07,SS08,SHSL09,SS10}). It is well known, that some types of Hamiltonian systems have a finite energy of escape. For lower values of the energy the equipotential surfaces of the systems are closed and therefore escape is impossible. For energy levels beyond the escape energy however, these surfaces open and exit channels emerge through which the particles can escape to infinity. The literature is replete with studies of such ``open" or ``leaking" Hamiltonian systems (e.g., \cite{BBS09,CKK93,KSCD99,LT11,NH01,SHA03,STN02,SCK95,SKCD95,SKCD96,Z14a,Z14b,Z15d}). Here it should emphasized, that all the above-mentioned references regarding previous studies in chaotic scattering and open Hamiltonian systems are exemplary rather than exhaustive, taking into account that a large quantity of related literature exists.

Usually, the infinity acts as an attractor for an escape particle, which may escape through different channels (exits) on the equipotential curve or on the equipotential surface depending whether the dynamical system has two or three degrees of freedom, respectively. Therefore, it is quite possible to obtain the basins of escape, similar to the basins of attraction in dissipative systems or even the Newton-Raphson fractal structures. Basins of escape have been studied in several papers (e.g., \cite{AVS01,BGOB88,C02,KY91,PCOG96}). The reader can find more details about the basins of escape in \cite{C02}.

The three-body problem is a paradigmatic case in celestial mechanics. This problem deals with the gravitationally interacting celestial bodies and predicts their Newtonian motion. All planets and asteroids in our Solar System move around the Sun, while in the same manner the moons orbit their host planets. The system Sun-planet-moon, or the Sun-planet-asteroid can be considered as typical examples of the three-body problem. In particular, the three-body problem of the Sun-planet-asteroid system can be significantly simplified since the mass of the asteroid is always negligible with respect to the masses of the Sun and the planet. This means that the gravitational influence of the asteroid on the Sun and also on the planet can be easily omitted from the equation describing the planetary system. Using this assumption then the three-body problem becomes the usual restricted three-body problem (RTBP) \cite{S67}. In this case there are two possibilities regarding the type of motion of the two primary bodies around their common center of mass: the circular RTBP and the elliptic RTBP.

Escaping and colliding orbits in the RTBP is another typical example. In \cite{N04,N05} and \cite{Z15g} a systematic orbit classification was carried out in the planar circular RTBP. In particular, initial conditions of orbits were classified into three main categories: (i) bounded; (ii) escaping and (iii) collision, while bounded regular orbits were further classified into orbital families regarding their motion around the two primary bodies. \cite{dAT14} investigated the orbital dynamics of the two dimensional version of the Earth-Moon system in a scattering region around the Moon, while in \cite{Z16} we the escape dynamics of the full three degrees of freedom system was unveiled. In the same vein, in \cite{Z15e} and \cite{Z15f} we classified orbits thus revealing the orbital structure of the Saturn-Titan and Pluto-Charon planetary systems, respectively.

In \cite{JLP03} the authors studied the motion of three masses in a plane interacting with a central potential proportional to $1/r^2$. In this paper we shall numerically investigate the nature of orbits in the planar circular restricted three-body problem (PCRTBP) where one of the two primaries has a strong gravitational field. In particular, we shall consider cases where the attracting force of one of the primaries is stronger with respect to the attracting force of the other primary modelled by a classical Newtonian potential. At this point we would like to mention the related problem of modified Newtonian gravity (also known as MOND) \cite{M83}. MOND however stands for corrections to the classical Newtonian  $1/r^2$ force in the form of a polynomial (Taylor) series, which are seem to be important as the scale length grows.

The structure of the paper is as follows: In Section \ref{mod} we present in detail the properties of the mathematical model. All the computational methods we used in order to obtain the classification of the orbits are described in Section \ref{cometh}. In the following Section, we conduct a thorough and systematic numerical investigation revealing the overall orbital structure of the PCRTBP by classifying orbits into categories. Our paper ends with Section \ref{disc} where the discussion and the conclusions of our research are given.

\section{Presentation of the model potential}
\label{mod}

It would be very informative to briefly recall the basic properties of the PCRTBP \cite{S67}. The two main bodies, called primaries move on circular orbits\footnote{In the Appendix we shall demonstrate that Keplerian circular orbits are also possible in the case where we have gravity stronger than the classical Newtonian one.} around their common center of gravity. The third body (also known as test particle) moves in the same plane under the gravitational field of the two primaries. It is assumed that the motion of the two primaries is not perturbed by the third body since the third body's mass is much smaller with respect to the masses of the two primaries.

The units of length, mass and time are taken so that the sum of the masses, the distance between the primaries and the angular velocity is unity, which sets the gravitational constant $G = 1$. A rotating rectangular system whose origin is the center of mass of the primaries and whose $Ox$-axis contains the primaries is used. The mass parameter is $\mu = m_2/(m_1 + m_2)$, where $m_1 = 1 - \mu$ and $m_2 = \mu$ are the dimensionless masses of the primaries with $m_1 > m_2$, such that $m_1 + m_2 = 1$, while their centers $P_1$ and $P_2$ of the primaries are located at $(-\mu, 0)$ and $(1-\mu,0)$, respectively.

The total time-independent effective potential function in the rotating frame of reference is
\begin{equation}
\Omega(x,y) = \frac{(1 - \mu)}{r_1} + \frac{\mu}{r_2^{p}} + \frac{1}{2}\left(x^2  + y^2 \right),
\label{pot}
\end{equation}
where
\[
r_1 = \sqrt{\left(x + \mu\right)^2 + y^2},
\]
\begin{equation}
r_2 = \sqrt{\left(x + \mu - 1\right)^2 + y^2},
\label{dist}
\end{equation}
are the distances to the respective primaries.

\begin{figure*}[!t]
\centering
\resizebox{\hsize}{!}{\includegraphics{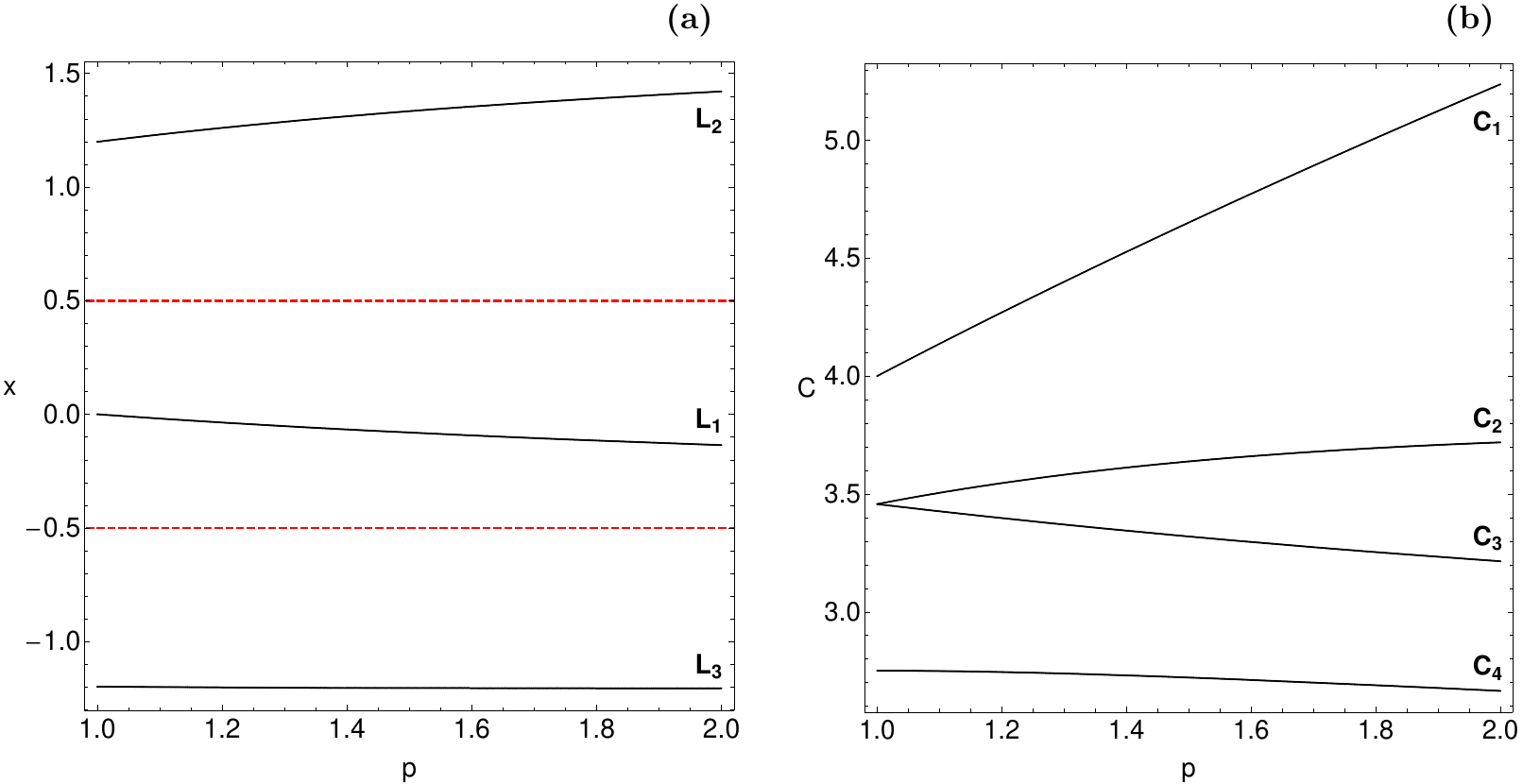}}
\caption{(a-left): Evolution of the position of the collinear Lagrange points $L_1$, $L_2$ and $L_3$ as a function of the power $p$ of the gravitational potential of the second primary. The horizontal red dashed lines indicate the position of the centers of the two primaries. (b-right): Evolution of the critical values of the Jacobi integral of motion as a function of the power $p$.}
\label{theor}
\end{figure*}

Looking at Eq. (\ref{pot}) we see that the gravitational potential of the first primary body is a classical Newtonian potential of the form $1/r$. The same applies for the second primary only when $p = 1$. In this paper we shall investigate the orbital dynamics of the test particle when $p \in [1, 2)$. When $p > 1$ we have the case of a stronger gravitational field with respect to the classical Newtonian gravity where $p = 1$. We choose $\mu = 1/2$ (which means that both primaries have equal masses) so that the only difference between the two primary bodies to be the factor of the gravitational potential. This choice will help us to unveil the influence of the gravity on the nature of orbits since the gravitational power $p$ will be the only variable parameter. When $p$ is larger than 1, then the interaction is stronger for short distances and is weaker for large distances. In our case the gravitational attraction of primary 2 is stronger in the scattering region.

The scaled equations of motion describing the motion of the third body in the corotating frame read
\[
\Omega_x = \frac{\partial \Omega}{\partial x} = \ddot{x} - 2\dot{y},
\]
\begin{equation}
\Omega_y = \frac{\partial \Omega}{\partial y} = \ddot{y} + 2\dot{x}.
\label{eqmot}
\end{equation}
The dynamical system (\ref{eqmot}) admits the well known Jacobi integral of motion
\begin{equation}
J(x,y,\dot{x},\dot{y}) = 2\Omega(x,y) - \left(\dot{x}^2 + \dot{y}^2 \right) = C,
\label{ham}
\end{equation}
where $\dot{x}$ and $\dot{y}$ are the velocities, while $C$ is the Jacobi constant which is conserved and defines a three-dimensional invariant manifold in the total four-dimensional phase space. Therefore, an orbit with a given value of its energy integral is restricted in its motion to regions in which $C \leq 2 \Omega(x,y)$, while all other regions are energetically forbidden to the third body. The energy value $E$ is related with the Jacobi constant by $C = - 2E$.

\begin{figure*}[!t]
\centering
\resizebox{\hsize}{!}{\includegraphics{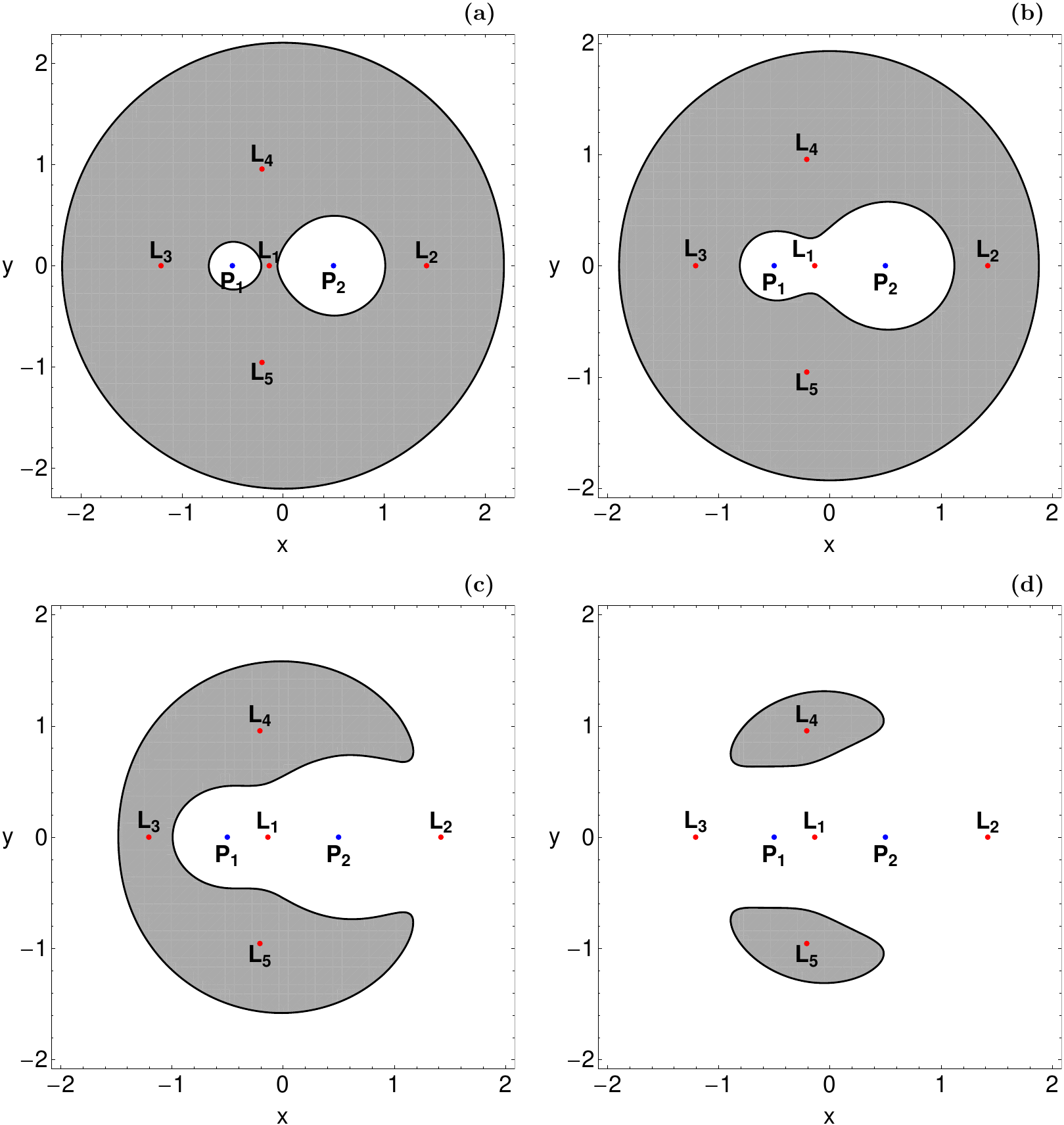}}
\caption{The first four Hill's regions configurations for the PCRTBP system with strong gravitational field, when $p = 1.95$. The white domains correspond to the Hill's regions, gray shaded domains indicate the energetically forbidden regions, while the thick black lines depict the Zero Velocity Curves (ZVCs). The red dots pinpoint the position of the Lagrange points, while the positions of the centers of the two primary bodies are indicated by blue dots. (a): $C = 5.5$; (b): $C = 4.47832$; (c): $C = 3.46658$; (d): $C = 2.9393$.}
\label{conts}
\end{figure*}

The dynamical system has five equilibrium points known as Lagrange points \cite{S67} at which
\begin{equation}
\frac{\partial \Omega}{\partial x} = \frac{\partial \Omega}{\partial y} = 0.
\label{lps}
\end{equation}
$L_1$, $L_2$, and $L_3$ (known as collinear points) are located on the $x$-axis, while the other two $L_4$ and $L_5$ are called triangular points and they are located in the vertices of equilateral triangles. It should be noted that the labeling of the first three Lagrange points is not consistent throughout the literature. In our case, we use the most popular labeling according to which $L_1$ is between the two primaries, $L_2$ is at the right side of $P_2$, while $L_3$ is at the left side of $P_1$.

The position of the five Lagrange points is actually a function of the power $p$ of the gravitational potential of the second primary. In Fig. \ref{theor}a we show how the position of the collinear points $L_1$, $L_2$ and $L_3$ evolves with respect to the value of the power $p$. The values of the Jacobi integral of motion at the Lagrange points $L_i, i = 1, ..., 5$ are denoted by $C_i$ and they are critical values (note that $C_4 = C_5$). Fig. \ref{theor}b shows the evolution of the $C_i$ critical values as a function of the power $p$. We see that the values of $C_1$, $C_2$ increase with increasing $p$, while on the other hand the values of $C_3$ and $C_4$ are reduced as the gravity becomes stronger.

The projection of the four-dimensional phase space onto the configuration (or position) space $(x,y)$ is called the Hill's regions and is divided into three domains: (i) the interior region for $x(L_3) \leq x \leq x(L_2)$; (ii) the exterior region for $x < x(L_3)$ and $x > x(L_2)$; (iii) the forbidden regions. The boundaries of these Hill's regions are called Zero Velocity Curves (ZVCs) because they are the locus in the configuration $(x,y)$ space where the kinetic energy vanishes. The structure of the Hill's regions strongly depends on the value of the Jacobi constant and also on the value of the power of the potential. There are five distinct cases regarding the Hill's regions:
\begin{itemize}
  \item $C > C_1$: All necks are closed, so there are only bounded and collision basins (see Fig. \ref{conts}a).
  \item $C_2 < C < C_1$: Only the neck around $L_1$ is open thus allowing orbits to move around both primaries (see Fig. \ref{conts}b).
  \item $C_3 < C < E_2$: The neck around $L_2$ opens, so orbits can enter the exterior region and escape form the system (see Fig. \ref{conts}c).
  \item $C_4 < C < C_3$: The necks around both $L_2$ and $L_3$ are open, therefore orbits are free to escape through two different escape channels (see Fig. \ref{conts}d).
  \item $C < C_4$: The banana-shaped forbidden regions disappear, so motion over the entire configuration $(x,y)$ space is possible.
\end{itemize}
In Fig. \ref{conts}(a-d) we present the structure of the first four possible Hill's region configurations for $p = 1.95$. We observe in Fig. \ref{conts}d the two openings (exit channels) near the Lagrange points $L_2$ and $L_3$ through which the test body can enter the exterior region and then leak out. In fact, we may say that these two exits act as hoses connecting the interior region of the system with the ``outside world" of the exterior region.

\section{Computational methods}
\label{cometh}

The motion of the massless test particle is restricted to a three-dimensional surface $C = const$, due to the existence of the Jacobi integral of motion. With polar coordinates $(r,\phi)$ in the center of the mass system of the corotating frame the condition $\dot{r} = 0$ defines a two-dimensional surface of section, with two disjoint parts $\dot{\phi} < 0$ and $\dot{\phi} > 0$. Each of these two parts has a unique projection onto the configuration $(x,y)$ space. In order to explore the orbital structure of the system we need to define sets of initial conditions of orbits whose properties will be identified. For this purpose, we define for several values of the Jacobi constant $C$, as well as for the gravitational power $p$ dense uniform grids of $1024 \times 1024$ initial conditions regularly distributed on the configuration $(x,y)$ space inside the area allowed by the value of the Jacobi constant. Following a typical approach, the orbits are launched with initial conditions inside a certain region, called scattering region, which in our case is a square grid with $-2\leq x,y \leq 2$.

In the PCRTBP system the configuration space extends to infinity thus making the identification of the type of motion of the test body for specific initial conditions a rather demanding task. There are three possible types of motion for the test particle: (i) bounded motion around one of the primaries, or even around both; (ii) escape to infinity; (iii) collision into one of the two primaries. Now we need to define appropriate numerical criteria for distinguishing between these three types of motion. The motion is considered as bounded if the test body stays confined for integration time $t_{\rm max}$ inside the system's disk with radius $R_d$ and center coinciding with the center of mass. Obviously, the higher the values of $t_{\rm max}$ and $R_d$ the more plausible becomes the definition of bounded motion and in the limit $t_{\rm max} \rightarrow \infty$ the definition is the precise description of bounded motion in a finite disk of radius $R_d$. Consequently, the higher these two values, the longer the numerical integration of initial conditions of orbits lasts. In our calculations we choose $t_{\rm max} = 10^4$ and $R_d = 10$ as in \cite{N04,N05} and \cite{Z15a,Z15b,Z15c}. We decided to include a relatively high disk radius $(R_d = 10)$ in order to be sure that the orbits will certainly escape from the system and not return back to the interior region\footnote{The most safe and efficient way to determine if an orbit escapes or not is the value of the total orbital energy of the particle measured by an observer in the inertial frame of reference. In particular, if the total orbital energy in the inertial frame is negative, the test particle might return back to the scattering region. On the contrary, if the total orbital energy becomes positive the test particle escapes, beyond any doubt, and it will never come back \cite{BTS96}. Our previous numerical experience (e.g., \cite{Z15a,Z15b,Z15c}) strongly suggests that the total orbital energy of the test-particle in the inertial frame becomes positive much sooner than it takes for the massless particle to cross the disk with radius $R_d = 10$. Thus we may claim that our escape criterion used in the previous series of papers, and also in the present one, is both correct and safe. In the following Section we will present numerical evidence proving the validity of our escape criterion.}. Moreover, an orbit is identified as escaping and the numerical integration stops if the test particle intersects the system's disk with velocity pointing outwards at a time $t_{\rm esc} < t_{\rm max}$. Finally, a collision with one of the primaries occurs if the test particle, assuming it is a point mass, crosses the disk with radius $R_{\rm col}$ around the primary. For both primaries we choose $R_{\rm col} = 10^{-4}$.

At this point, we would like to point out that the necessity to consider much larger radii $(R_d > 10)$ to define whether a test particle is escaping or not, is tantamount to be able to approximate, for very high distances, the time-dependent long range potential $1/r$ as a static one. Only in that case it is possible to argue that the energy (which is not constant in this case) has very small fluctuations and then becomes a good parameter to speak about the fate of the orbits. Yet, the orbits that do reach such a large value of $R_d$ may still have asymptotically negative, positive, or even zero total orbital energy. Scattering orbits must have at least zero asymptotic energy (i.e., display asymptotic parabolic motion). Restricting the value of the radius of the disk to 10 is equivalent to cut-off the long range potential artificially, and then modify the orbital structure of the phase space. However in this work we deal only with the frozen time-independent potential and therefore the value 10, regarding the radius of the disk which distinguishes between escaping and non-escaping or trapped motion, is sufficient for the desired computations.

As it was stated earlier, in our computations, we set $10^4$ time units as a maximum time of numerical integration. The vast majority of escaping orbits (regular and chaotic) however, need considerable less time to escape from the system (obviously, the numerical integration is effectively ended when an orbit moves outside the system's disk and escapes). Nevertheless, we decided to use such a vast integration time just to be sure that all orbits have enough time in order to escape. Remember, that there are the so called ``sticky orbits" which behave as regular ones during long periods of time. Here we should clarify, that orbits which do not escape after a numerical integration of $10^4$ time units are considered as non-escaping or trapped.

The equations of motion (\ref{eqmot}) for the initial conditions of all orbits are forwarded integrated using a double precision Bulirsch-Stoer \verb!FORTRAN 77! algorithm (e.g., \cite{PTVF92}). Here we should emphasize, that our previous numerical experience suggests that the Bulirsch-Stoer integrator is both faster and more accurate than a double precision Runge-Kutta-Fehlberg algorithm of order 7 with Cash-Karp coefficients. Throughout all our computations, the Jacobi integral of motion of Eq. (\ref{ham}) was conserved better than one part in $10^{-11}$, although for most orbits it was better than one part in $10^{-12}$. For collision orbits where the test body moves inside a region of radius $10^{-2}$ around one of the primaries the Lemaitre's global regularization method is applied \cite{S67}. All graphical illustrations presented in this work have been created using version 10.3 of the software Mathematica$^{\circledR}$ \cite{W03}.

\section{Orbit classification}
\label{orbclas}

The main numerical task is to classify initial conditions of orbits in the $\dot{\phi} < 0$ part\footnote{We choose the $\dot{\phi} < 0$ instead of the $\dot{\phi} > 0$ part simply because in \cite{Z15a} we seen that it contains more interesting orbital content.} of the surface of section $\dot{r} = 0$ into three categories: (i) bounded orbits; (ii) escaping orbits and (iii) collision orbits. Moreover, two additional properties of the orbits will be examined: (i) the time-scale of collision and (ii) the time-scale of the escapes (we shall also use the terms escape/collision period or escape/collision rates). In this work we shall explore these dynamical quantities for various values of the total orbital energy, as well as for the power $p$ of the gravitational potential. In particular, we choose four energy levels which correspond to the last four Hill's regions configurations. The first Hill's regions configurations contain only bounded and collision orbits around the two primaries, so the orbital content is not so interesting.

In the following color-coded grids (or orbit type diagrams - OTDs) each pixel is assigned a color according to the orbit type. Thus the initial conditions of orbits on the $(x,y)$-plane are classified into bounded orbits, unbounded or escaping orbits and collision orbits. In this special type of Poincar\'{e} Surface of Section (PSS) the phase space emerges as a closed and compact mix of escape basins, collision basins and stability regions. Our numerical calculations indicate that apart from the escaping and collision orbits there is also a considerable amount of non-escaping orbits. In general terms, the majority of non-escaping regions corresponds to initial conditions of regular orbits, where an adelphic integral of motion is present, restricting their accessible phase space and therefore hinders their escape.

In the following we are going to explore the orbital content of the configuration $(x,y)$ space in four different energy cases regarding the value of the Jacobi constant. In every case we choose four values of the power $p$ of the gravitational potential always in the interval $[1,2)$. For every value of $p$ the energy level is different. However, the chosen energy levels corresponding to the last four Hill' regions configurations admit the following relations: $(C_1 + C_2)/2$, $(C_2 + C_3)/2$, $(C_3 + C_4)/2$, $C_4$.

\subsection{Energy region I: $C_1 \geq C > C_2$}
\label{ss1}

\begin{figure*}[!t]
\centering
\resizebox{\hsize}{!}{\includegraphics{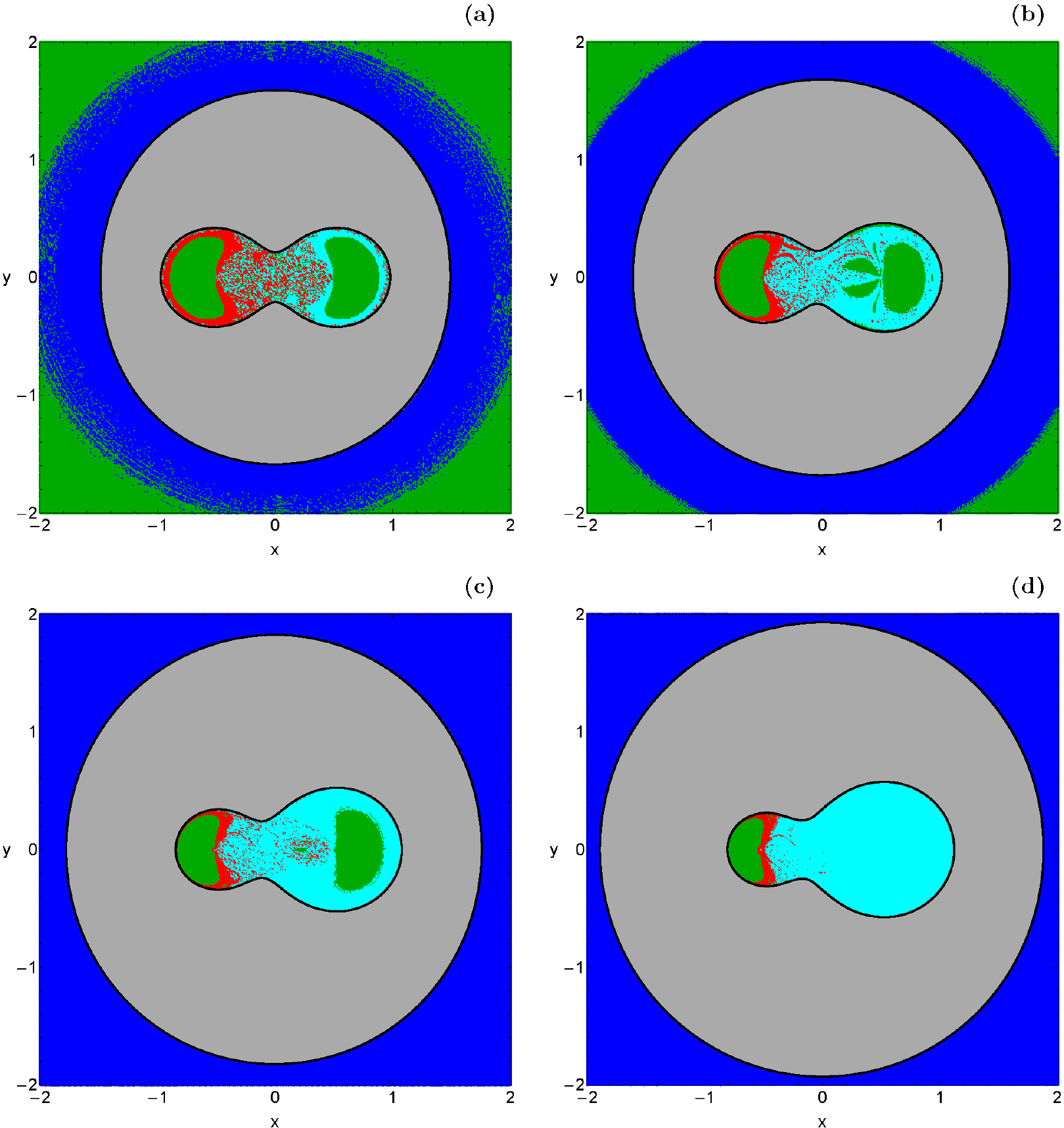}}
\caption{The orbital structure of the $\dot{\phi} < 0$ part of the surface of section $\dot{r} = 0$ when: (a-upper left): $p = 1$, $C = 3.72839$; (b-upper right): $p = 1.2$, $C = 3.90788$; (c-lower left): $p = 1.6$, $C = 4.21647$; (d-lower right): $p = 1.95$, $C = 4.44754$. The color code is the following: bounded orbits (green), collision orbits to primary 1 (red), collision orbits to primary 2 (cyan), and  escaping orbits (blue).}
\label{hr1}
\end{figure*}

In this energy region all transport channels are closed and therefore inside the interior region there is only bounded and collision motion. In Fig. \ref{hr1}(a-d) the OTD decompositions of the $\dot{\phi} < 0$ part of the surface of section $\dot{r} = 0$ reveal the orbital structure of the configuration $(x,y)$ space for four values of the power $p$ of the gravitational potential. The black solid lines denote the ZVC, while the inaccessible forbidden regions are marked in gray. The color of a point represents the orbit type of a test particle which has been launched with pericenter position at $(x,y)$. In every panel the value of the Jacobi constant is $(C_1 + C_2)/2$, where in each case the critical values of the Jacobi integral of motion correspond to the particular value of $p$. In Fig. \ref{hr1}a, where $p = 1$, we see that the OTD is symmetric with respect to both axes ($x$ and $y$). In the case of the classical Newtonian gravity $(p = 1)$ bounded motion is present in both the interior and the exterior region. In particular, two stability islands are present near the centers of the two primaries. These islands are composed of initial conditions of orbits which circulate around one of the primary bodies. In the exterior region the stability region has a shape of an annulus and corresponds to orbits that circulate around both primaries. Around the two stability islands in the interior region collision basins are present, while the central region of the OTD contains a highly fractal\footnote{When we state that an area is fractal we simply mean that it has a fractal-like geometry without conducting any specific calculations as in \cite{AVS09}.} mixture of collision orbits. Escaping orbits do exist, only in the exterior region, forming a ring-shape escape basin just outside the forbidden region. Panels (b)-(d) of Fig. \ref{hr1} correspond to the cases of stronger gravitational filed of primary 2. In panel (b) we have the case where $p = 1.2$. One may see that even a slight increase of the gravitational filed of the second primary leads to changes regarding the orbital structure of the configuration space. The changes are more significant in the interior region, where the collision basin to primary 2 is larger, while the extent of the collision basins to primary 1 has been reduced. The shape of the stability region near primary 2 has also been changed. With a much closer look we observe that the boundaries between the escape basin and the stability annulus located in the exterior region are more smooth with respect to that observed in panel (a). In panel (c) the gravitational field of primary 2 is stronger since $p = 1.6$. It is seen that the degree of fractality of the interior region has been reduced. At the same time the area of the energetically forbidden regions has been increased. This is actually the reason of why the outer stability island is not visible now (additional numerical calculations suggest that it is still present however it is located at larger distance form the center). The case where $p = 1.95$ is presented in panel (d) of Fig. \ref{hr1}. Here things are very different since the collision basin to primary 2 dominates the interior region, while the stability island near center $P_2$ has disappeared. On the other hand we did not observe any differences in the exterior region. Our results suggest that in this energy region the increase of the power $p$ of the potential affects mostly the region around primary 2, while the orbital structure around primary 1 is much less influenced by the increase of the gravitational field.

\begin{figure*}[!t]
\centering
\resizebox{\hsize}{!}{\includegraphics{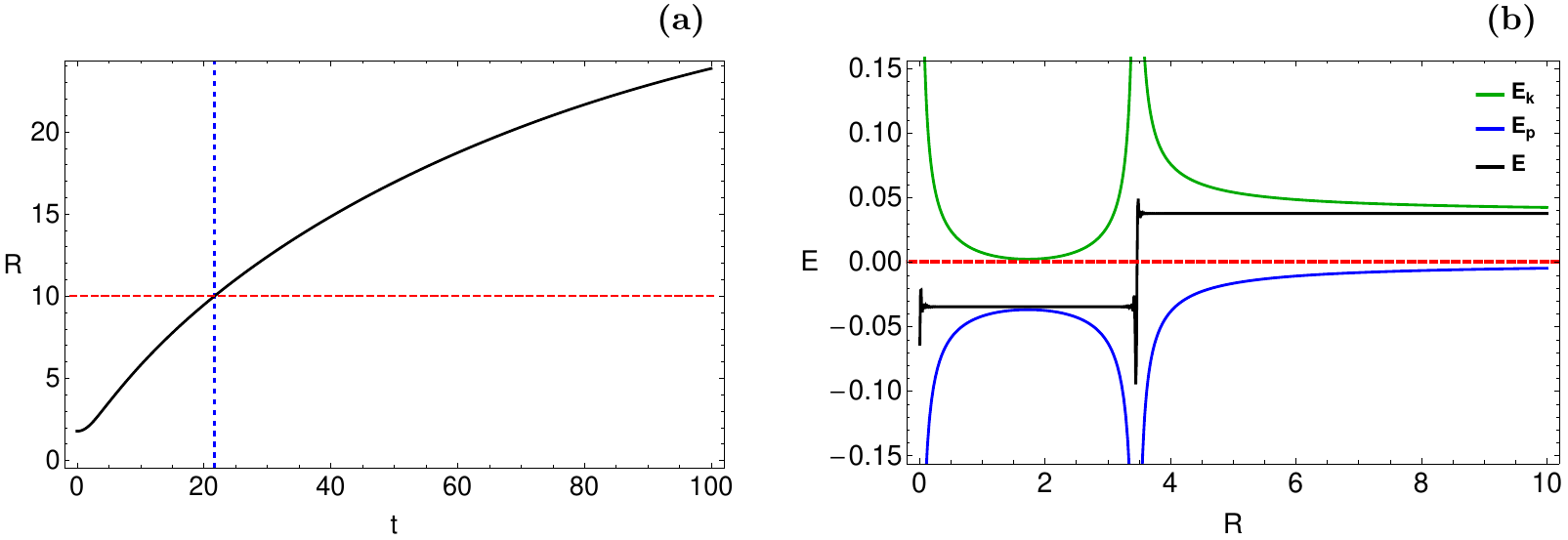}}
\caption{(a-left): Evolution of distance $R$ of the origin as a function of time. The horizontal red dashed line indicates the escape threshold radius $R_d = 10$, while the vertical blue dashed line denotes the escape time of the orbit. (b-right): Time development of the kinetic energy $E_k$ (green), the potential energy $E_p$ (red), and the total orbital energy $E$ (black) of the test particle measured by an observer in the inertial frame of reference. The energy level $E = 0$, i.e. the threshold for escape to infinity, is marked as red, dashed line. The initial conditions of the orbit and more details are given in the text.}
\label{orb}
\end{figure*}

\begin{figure}[!t]
\begin{center}
\includegraphics[width=\hsize]{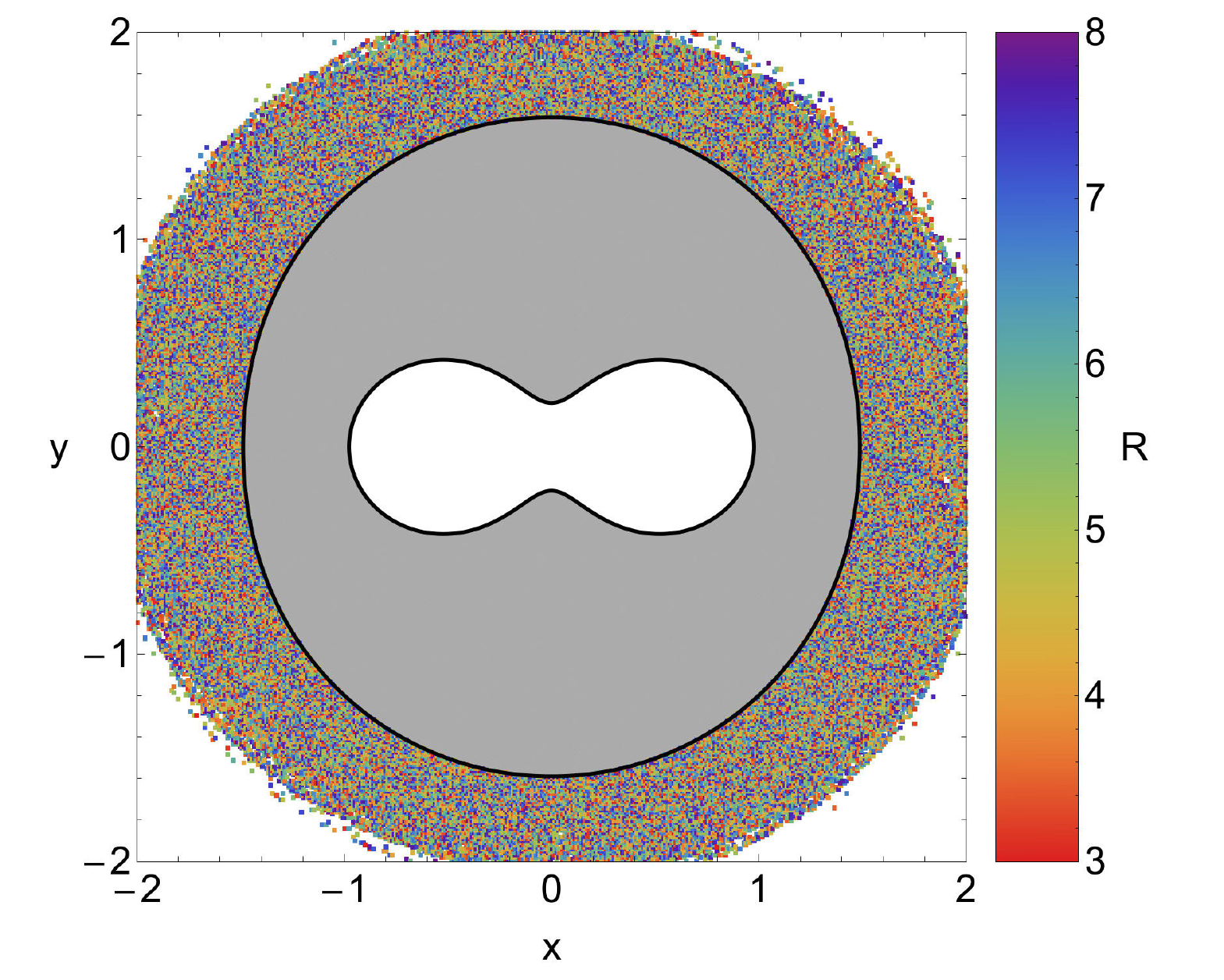}
\end{center}
\caption{Distribution of the radii $R$ at which the total orbital energy $E$ in the inertial frame of reference becomes positive, for $C = 3.72839$. The initial conditions of the orbits are the same as in Fig. \ref{hr1}a. Note that for all orbits $E$ becomes positive at lower radii than $R_d$.}
\label{Rds}
\end{figure}

\begin{figure*}[!t]
\centering
\resizebox{\hsize}{!}{\includegraphics{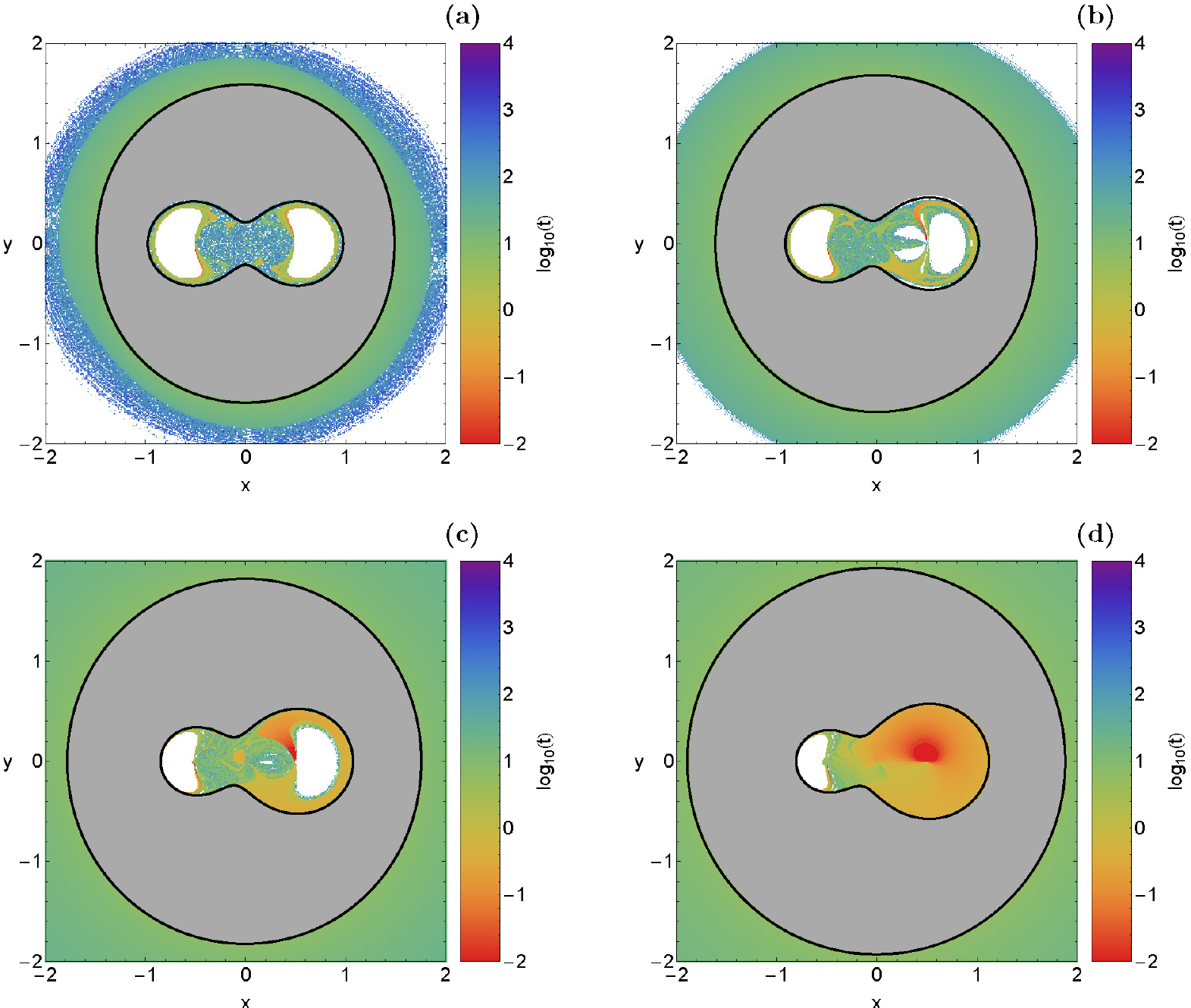}}
\caption{Distribution of the escape and collision time of the orbits on the $\dot{\phi} < 0$ part of the surface of section $\dot{r} = 0$ for the cases of Fig. \ref{hr1}. The bluer the color, the larger the escape/collision time. Initial conditions of bounded orbits are shown in white.}
\label{hr1t}
\end{figure*}

\begin{figure*}[!t]
\centering
\resizebox{\hsize}{!}{\includegraphics{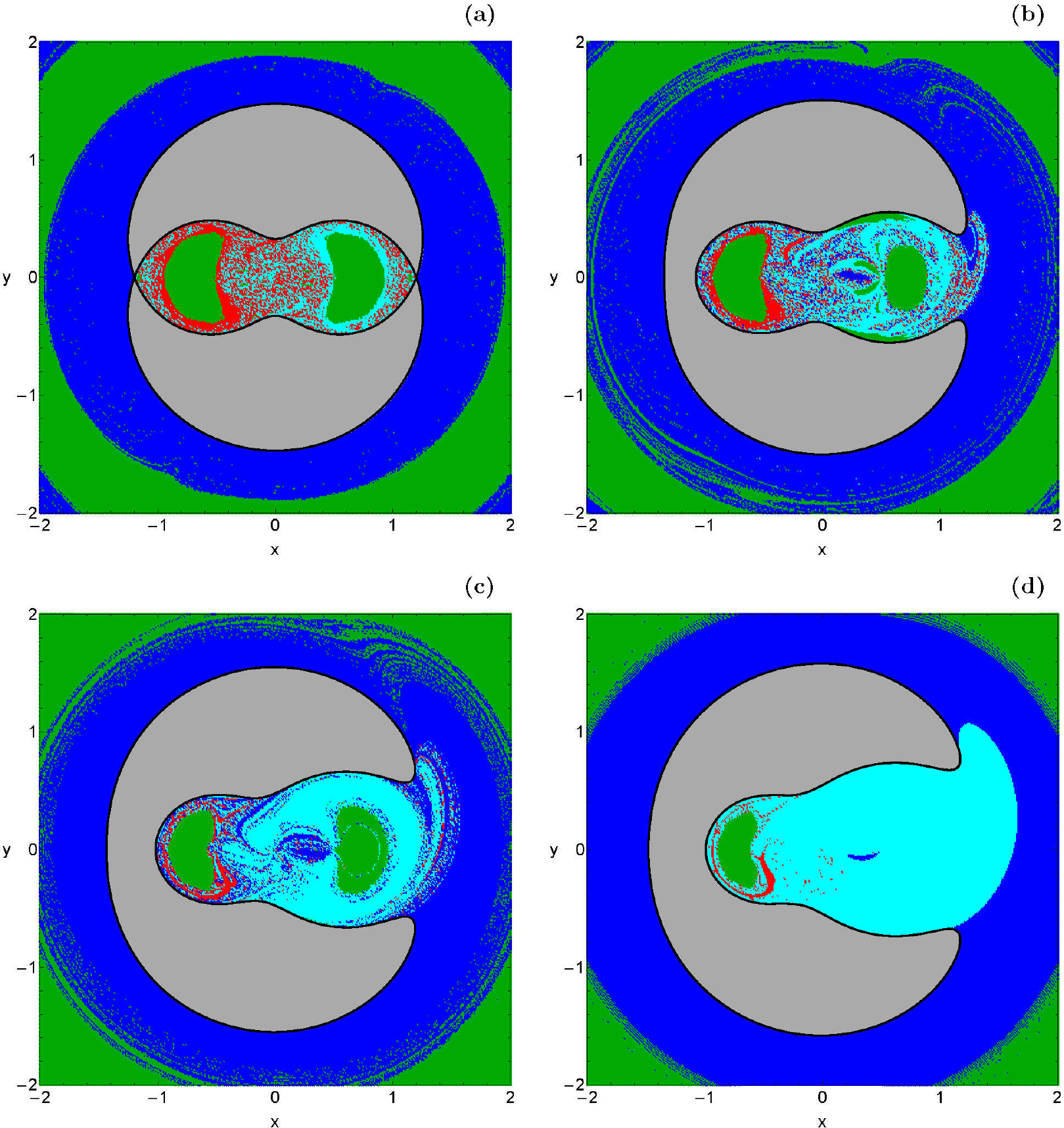}}
\caption{The orbital structure of the $\dot{\phi} < 0$ part of the surface of section $\dot{r} = 0$ when: (a-upper left): $p = 1$, $C = 3.45679$; (b-upper right): $p = 1.2$, $C = 3.47168$; (c-lower left): $p = 1.6$, $C = 3.47812$; (d-lower right): $p = 1.95$, $C = 3.46863$. The color code is the same as in Fig. \ref{hr1}.}
\label{hr2}
\end{figure*}

\begin{figure*}[!t]
\centering
\resizebox{\hsize}{!}{\includegraphics{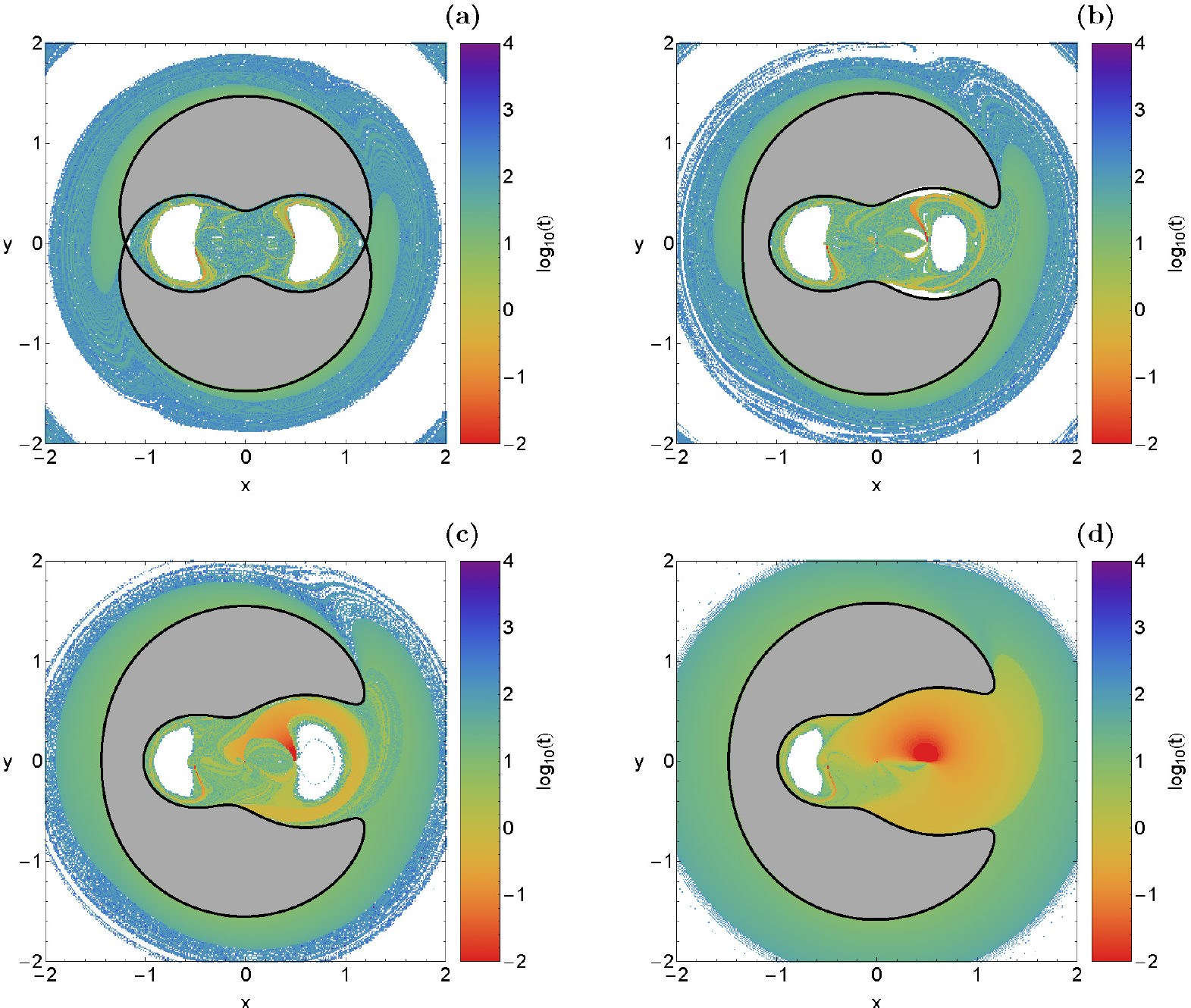}}
\caption{Distribution of the escape and collision time of the orbits on the $\dot{\phi} < 0$ part of the surface of section $\dot{r} = 0$ for the cases of Fig. \ref{hr1}.}
\label{hr2t}
\end{figure*}

\begin{figure*}[!t]
\centering
\resizebox{\hsize}{!}{\includegraphics{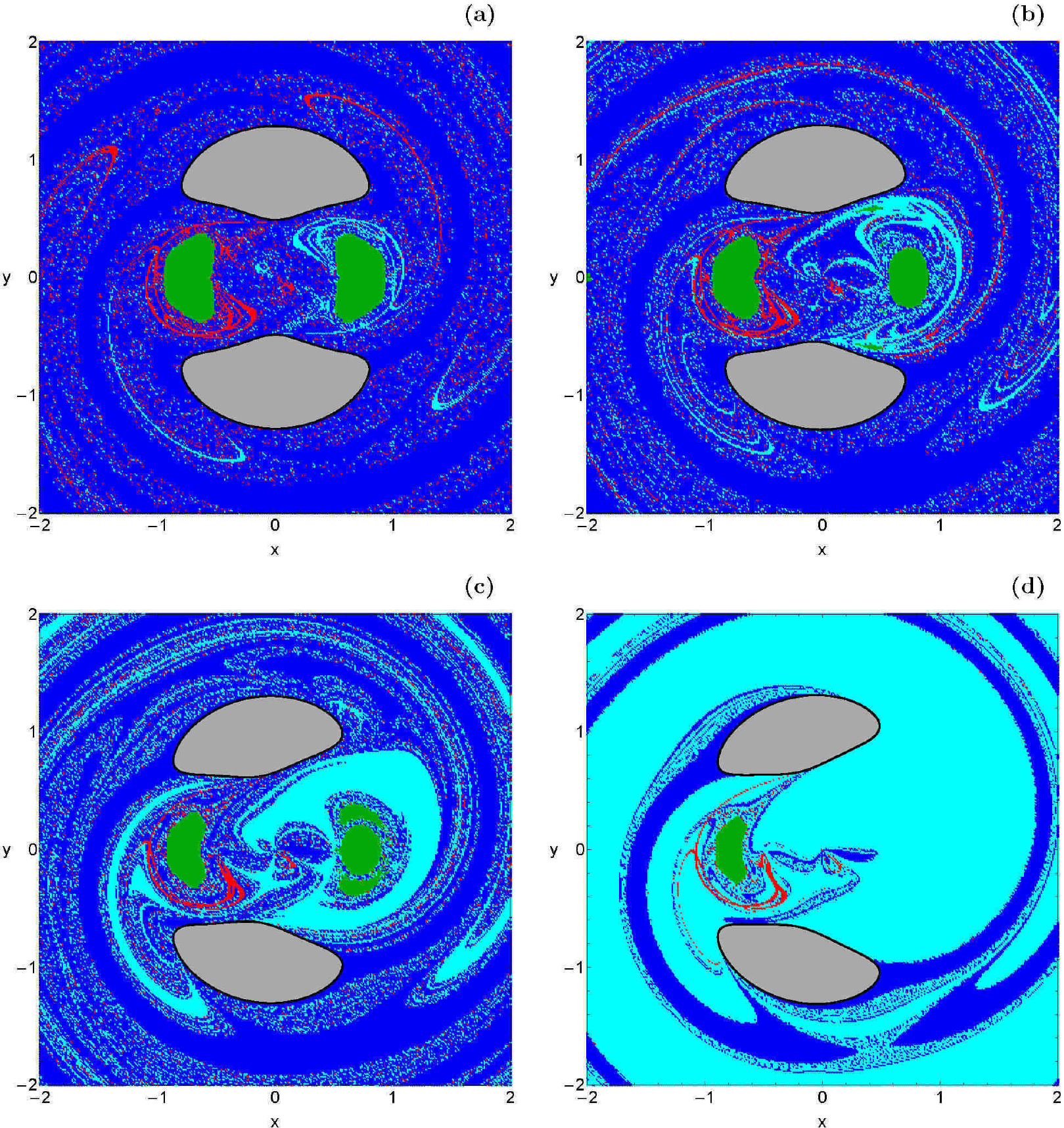}}
\caption{The orbital structure of the $\dot{\phi} < 0$ part of the surface of section $\dot{r} = 0$ when: (a-upper left): $p = 1$, $C = 3.10339$; (b-upper right): $p = 1.2$, $C = 3.07091$; (c-lower left): $p = 1.6$, $C = 3.00348$; (d-lower right): $p = 1.95$, $C = 2.94703$. The color code is the same as in Fig. \ref{hr1}.}
\label{hr3}
\end{figure*}

\begin{figure*}[!t]
\centering
\resizebox{\hsize}{!}{\includegraphics{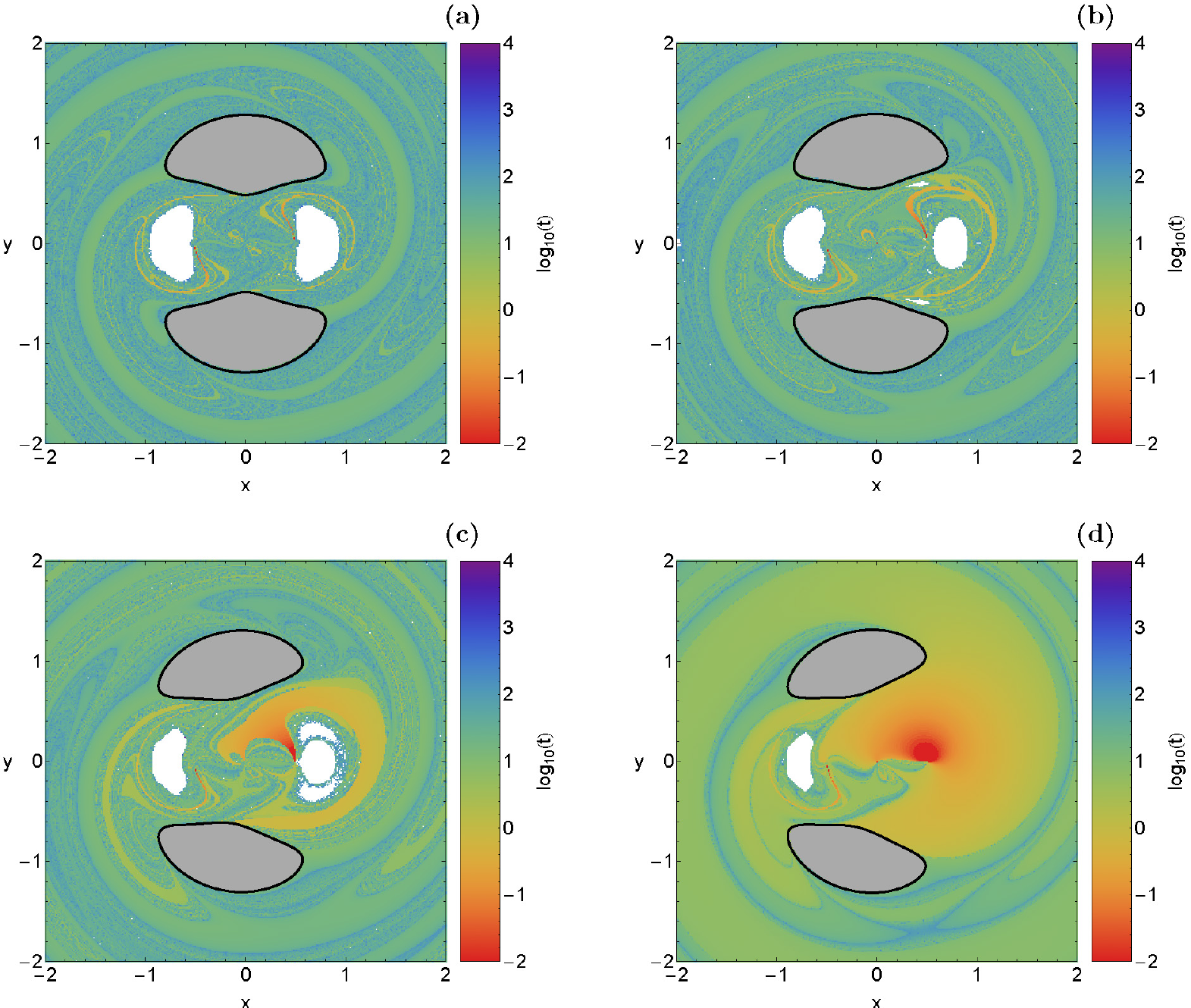}}
\caption{Distribution of the escape and collision time of the orbits on the $\dot{\phi} < 0$ part of the surface of section $\dot{r} = 0$ for the cases of Fig. \ref{hr1}.}
\label{hr3t}
\end{figure*}

\begin{figure*}[!t]
\centering
\resizebox{\hsize}{!}{\includegraphics{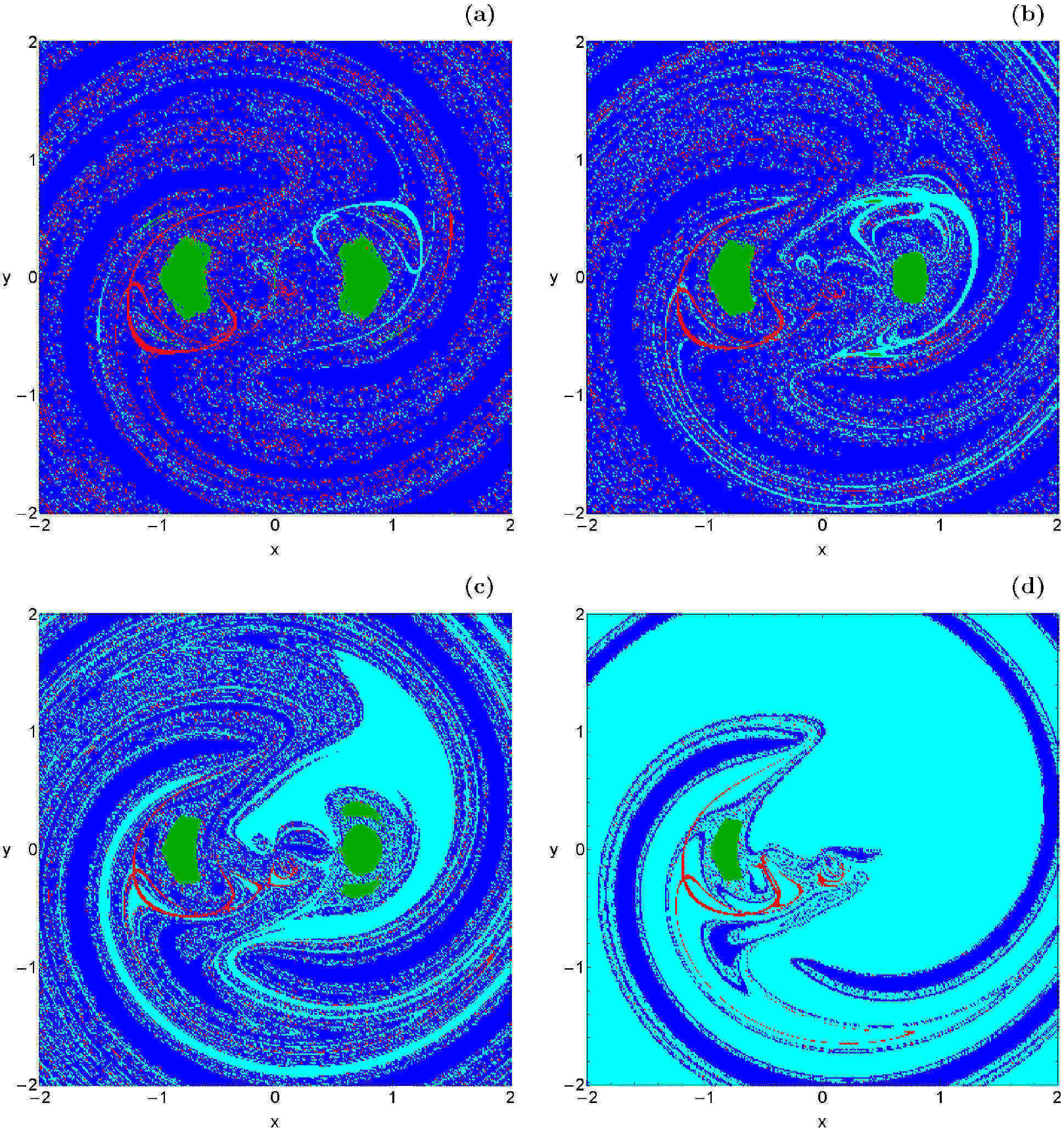}}
\caption{The orbital structure of the $\dot{\phi} < 0$ part of the surface of section $\dot{r} = 0$ when: (a-upper left): $p = 1$, $C = 2.75$; (b-upper right): $p = 1.2$, $C = 2.74426$; (c-lower left): $p = 1.6$, $C = 2.71067$; (d-lower right): $p = 1.95$, $C = 2.67031$. The color code is the same as in Fig. \ref{hr1}.}
\label{hr4}
\end{figure*}

\begin{figure*}[!t]
\centering
\resizebox{\hsize}{!}{\includegraphics{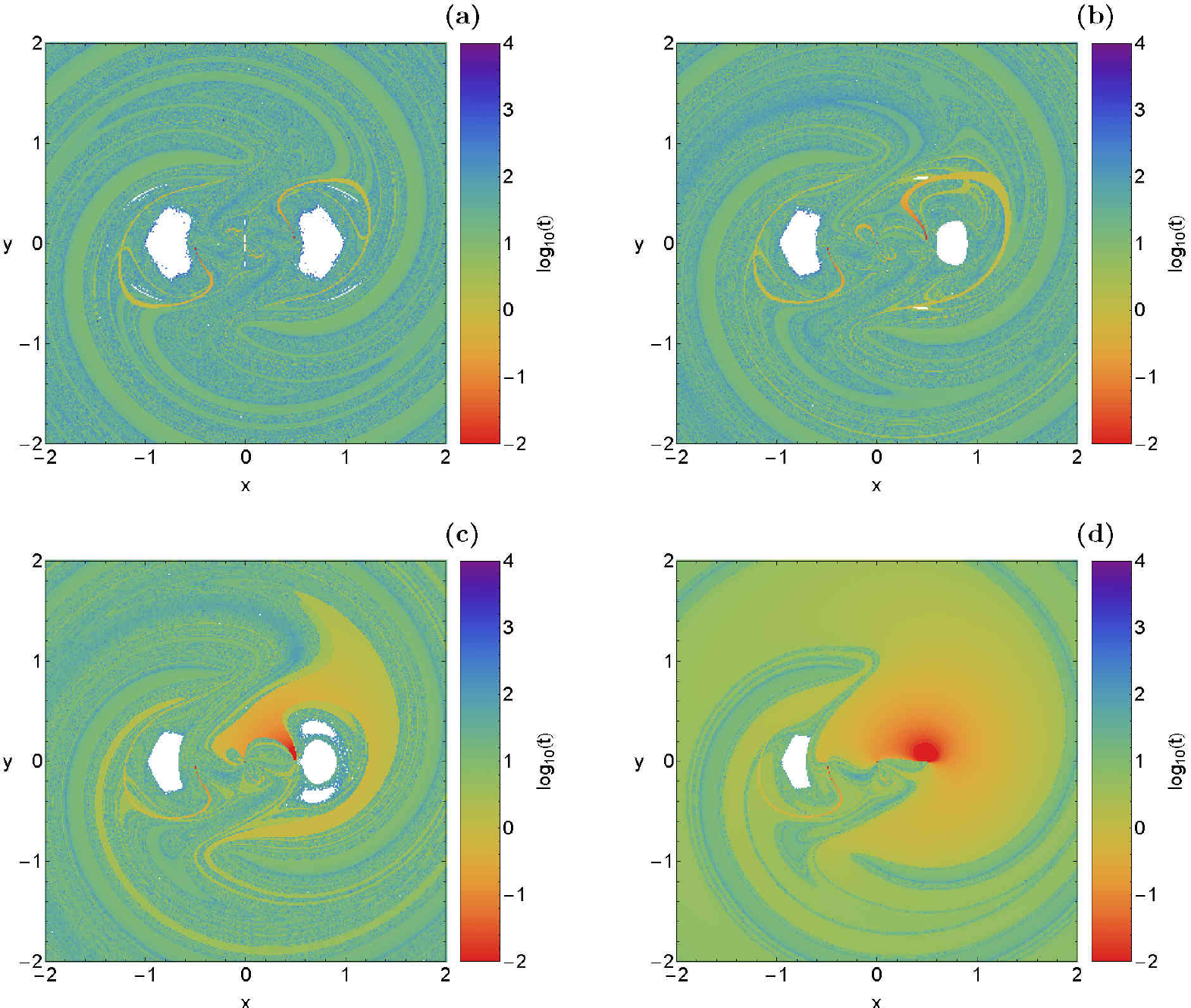}}
\caption{Distribution of the escape and collision time of the orbits on the $\dot{\phi} < 0$ part of the surface of section $\dot{r} = 0$ for the cases of Fig. \ref{hr1}.}
\label{hr4t}
\end{figure*}

In panel (a) of Fig. \ref{hr1} we see that the stability island located in the exterior region which is composed of initial conditions of orbits that circulate around both primaries bounds the inner escape basin. At first glance this behavior seems quite odd and one may misinterpret this phenomenon. However, the fact that the stability island of bounded motion is located outside the escape basin does not mean, by no means, that the orbits do not escape or that the escape radius $R_d$ is too small. It should be emphasized that similar orbital structures (escape regions bounded by stability islands) were also found in all earlier similar works (e.g., \cite{N04,N05,Z15a,Z15b,Z15c}). Remember that the color-coded OTDs on the configuration $(x,y)$ plane are just two-dimensional projections of the entire four-dimensional phase space. In this four-dimensional space, where all orbits live and move, no stability island bound their motion and therefore they are free to escape.

In order to prove that the orbits do escape we choose for $C = 3.72839$ an orbit with initial conditions $x_0 = -1.7$, $y_0 = 0.5$, that is inside the escape basin. We numerically integrate this orbit and we record its distance $R = \sqrt{x^2 + y^2}$ form the origin. In panel (a) of Fig. \ref{orb} we present the evolution of $R$ as a function of time. It is seen that about $t = 21.7$ time units the orbit crosses the escape threshold radius $R_d = 10$. Moreover, the value of the radius $R$ continues to grow with increasing time. Now the interesting question is: Will this orbit return back in the future? Or in other words, was the classification safe using the particular escape radius? To answer this question we computed the total orbital energy $E$ of the particle measured by an observer in the inertial frame of reference. Our results are given in panel (b) of Fig. \ref{orb} where we plotted in black the evolution of $E$ as a function of the radius $R$. We observe that the total orbital energy becomes positive at approximately $R = 3.47$, that is much lower than $R_d = 10$. Nevertheless, we cannot be sure yet whether our criterion is safe or not. To determine this we illustrate in Fig. \ref{Rds} a color-coded diagram containing all the initial conditions of escaping orbits of Fig. \ref{hr1}a. In this type of plot each initial condition is given a color according to the radius $R$ at which the total orbital energy $E$ in the inertial frame of reference becomes positive. Now it is more than evident that all orbits have already escaped at $R_d = 10$. In fact, after carrying our additional calculations (not given here) for other energy levels as well as for values of $p$ we did not find any orbit for which the total orbital energy $E$ to become positive at greater radius than $R_d$. Thus we may claim now that our escape threshold $R_d$ is both valid and safe.

In the following Fig. \ref{hr1t} we show how the escape and collision times of orbits are distributed on the configuration $(x,y)$ space for the four cases discussed in Fig. \ref{hr1}(a-d). Light reddish colors correspond to fast escaping/collionalal orbits, dark blue/purple colors indicate large escape/collion times, while white color denote stability islands of bounded motion. Note that the scale on the color bar is logarithmic. Inspecting the spatial distribution of various different ranges of escape time, we are able to associate medium escape time with the stable manifold of a non-attracting chaotic invariant set, which is spread out throughout this region of the chaotic sea, while the largest escape time values on the other hand, are linked with sticky motion around the stability islands of the two primary bodies. As for the collision time we see that for $p > 1$ a portion of orbits with initial conditions very close to the vicinity of the center of primary 2 collide with it almost immediately, within the first time steps of the numerical integration. As the value of $p$ increases and the gravitational filed of primary 2 becomes stronger the amount of these super fast collision orbits substantially grow.

\subsection{Energy region II: $C_2 \geq C > C_3$}
\label{ss2}

Our exploration is continued with the second energy region where the neck around $L_2$ opens thus allowing orbits with initial conditions in the interior region to enter the exterior region. The orbital structure of the configuration $(x,y)$ space is unveiled in Fig. \ref{hr2}(a-d) through the OTD decompositions of the $\dot{\phi} < 0$ part of the surface of section $\dot{r} = 0$. The four panels correspond to the same four values of $p$ used earlier in Fig. \ref{hr1}(a-d). In this case the values of the Jacobi are given by the formula $(C_2 + C_3)/2$. However when $p = 1$ due to the symmetry of the PCRTBP $C_2 = C_3$ thus in panel (a) the transport channel around $L_2$ is still closed since $C = C_2$. In general terms we may say that the orbital structure of the OTD for $p = 1$ is very similar to that observed earlier in Fig. \ref{hr1}a. Here the stability annulus in the exterior region is more clear since its boundaries are visible at the corners of the OTD. When $p = 1.2$ it is seen in panel (b) that the symmetry of the OTD is destroyed. Initial conditions of both types of collision orbits leak out through $L_2$, while at the same time initial conditions of escaping orbits enter the interior region. It should also be noted that the boundaries between the escape basin and the stability annulus in the exterior region are not so smooth as in panel (a). As the gravitational field of primary 2 becomes stronger one may observe in panel (c) that when $p = 1.6$ the orbital content around primary 1 starts to change. In particular the collision basin to primary 1 seems to become weaker, while the collision basin to primary 2 expands in the interior region. When $p = 1.95$ initial conditions of orbits that collide with primary 2 is the most populated type of orbits in the interior region, while only a tiny portion of escaping orbits are present in the interior region. Furthermore, it is evident that the extent of the collision basin around primary 1 has been further reduced. Once more, as in the previous energy case, the stability island around primary 2 disappears when the gravitational filed of the same primary is too strong $(p = 1.95)$, while the stability island around primary 1 seems to be unperturbed. In addition the boundaries between the stability annulus and the escape basin in the exterior region are fairly smooth again. Our calculations reveal that when $p = 1.95$ only initial conditions of collision orbits to primary 2 leak out through Lagrange point $L_2$, while on the other hand initial conditions of collision orbits to primary 1 are mainly located around the same primary.

The distribution of the escape and collision times of orbits on the configuration space is shown in Fig. \ref{hr2t}(a-d). One may observe that the results are very similar to those presented earlier in Fig. \ref{hr1t}, where we found that orbits with initial conditions inside the escape and collision basins have the smallest escape/collision rates, while on the other hand, the longest escape/collision rates correspond to orbits with initial conditions in the fractal regions of the OTDs. An interesting phenomenon is related with the escape rates of orbits with initial conditions in the exterior region. It is seen that the escape basin around the forbidden regions is divided into two parts: the inner shell with initial conditions of orbits that escape after about 10 time units and the outer shell composed of initial conditions of orbits with escape times larger than 100 time units.

\subsection{Energy region III: $C_3 \geq C > C_4$}
\label{ss3}

In the next energy region there are two channels (around $L_2$ and $L_3$) through which orbits with initial conditions inside the interior region can escape from the system. Fig. \ref{hr3}(a-d) presents the orbital structure of the configuration space through the OTD decompositions of the $\dot{\phi} < 0$ part of the surface of section $\dot{r} = 0$. In every panel the value of the Jacobi constant is $(C_3 + C_4)/2$, where in each case the critical values of the Jacobi integral of motion correspond to the particular value of $p$. In panel (a) which corresponds to $p = 1$ (classical Newtonian gravity) we observe that initial conditions of collision orbits to both primaries form thin spiral bands in the configuration $(x,y)$ space. However the vast majority of the OTD is dominated by initial conditions of orbits which escape from the system. Bounded motion is still present due to the existence of two stability islands near the centers of the primaries. In this energy range there are no bounded orbits which circulate around both primaries since the corresponding stability annulus in the exterior region is now absent. In panels (b) and (c) where $p$ is 1.2 and 1.6, respectively it is seen that the amount of orbits which collide with primary 2 grows forming well-defined basins of collision. On the other hand the rate of collision orbits to primary 1 constantly decreases. When $p = 1.95$ one may observe in panel (d) that initial conditions of collision orbits to primary 2 have taken over the configuration space, while the amount of escaping orbits has been significantly reduced and the stability island near $P_2$ has disappeared. In the same vein, initial conditions of orbits which collide with primary 1 are only present around the same primary thus forming a thin collision basin. Therefore we may conclude that in this energy region the power of the gravitational potential affects the entire configuration space.

In Fig. \ref{hr3t}(a-d) we depict the distribution of the escape and collision times of orbits on the configuration space. One can see similar outcomes with that presented in the two previous subsections. At this point, we would like to emphasize that the basins of escape can be easily distinguished in Fig. \ref{hr3t}, being the regions with intermediate colors indicating fast escaping orbits. Indeed, our numerical computations suggest that orbits with initial conditions inside these basins need no more than 10 time units in order to escape from the system. Furthermore, the collision basins are shown with reddish colors where the corresponding collision time is less than one time unit of numerical integration.

\subsection{Energy region IV: $C \leq C_4$}
\label{ss4}

The last energy case under consideration corresponds to the Hill's regions configuration in which the energetically forbidden regions disappear and the test particle has access to the entire configuration $(x,y)$ space. In Fig. \ref{hr4}(a-d) we present the orbital structure of the configuration space through the OTD decompositions of the $\dot{\phi} < 0$ part of the surface of section $\dot{r} = 0$. In every panel the energy is equal to $C_4$, known also as Trojan energy \cite{N05}, according to the corresponding value of $p$. In panel (a) where $p = 1$ we see that all types of motion are present thus creating several bounded, collision and escape basins. The last two types of basins have the shape of spirals around the center of the frame of reference. As the value of $p$ increases and the gravity of primary 2 becomes stronger several changes take place in the $(x,y)$ plane. The most important ones are the following: (i) The collision basins to primary 2 grows; (ii) the extent of the escape basins is reduced; (iii) the degree of fractality decreases. In panel (d) where $p = 1.95$ it is seen that the stability island composed of initial conditions of bounded orbits around primary 2 is no longer present, while the collision basins to the same primary dominate the configuration space. The stability island around primary 1 and the corresponding collision basins seem almost unperturbed by the increase of the gravitational field of primary 2. The distribution of the corresponding escape and collision times of orbits on the configuration space are presented in Fig. \ref{hr4t}(a-d). Once more the results are very similar to those presented in the previous subsections.

The OTDs shown in Figs. \ref{hr1}, \ref{hr2}, \ref{hr3} and \ref{hr4} have both fractal and non-fractal (smooth) boundary regions which separate the escape basins from the collision basins. Such fractal basin boundaries is a common phenomenon in leaking Hamiltonian systems (e.g., \cite{BGOB88,dMG02,dML99,STN02,ST03,TSPT04}). In the PCRTBP system the leakages are defined by both escape and collision conditions thus resulting in three exit modes. However, due to the high complexity of the basin boundaries, it is very difficult, or even impossible, to predict in these regions whether the test particle collides with one of the primary bodies or escapes from the dynamical system.

\begin{figure*}[!t]
\centering
\resizebox{\hsize}{!}{\includegraphics{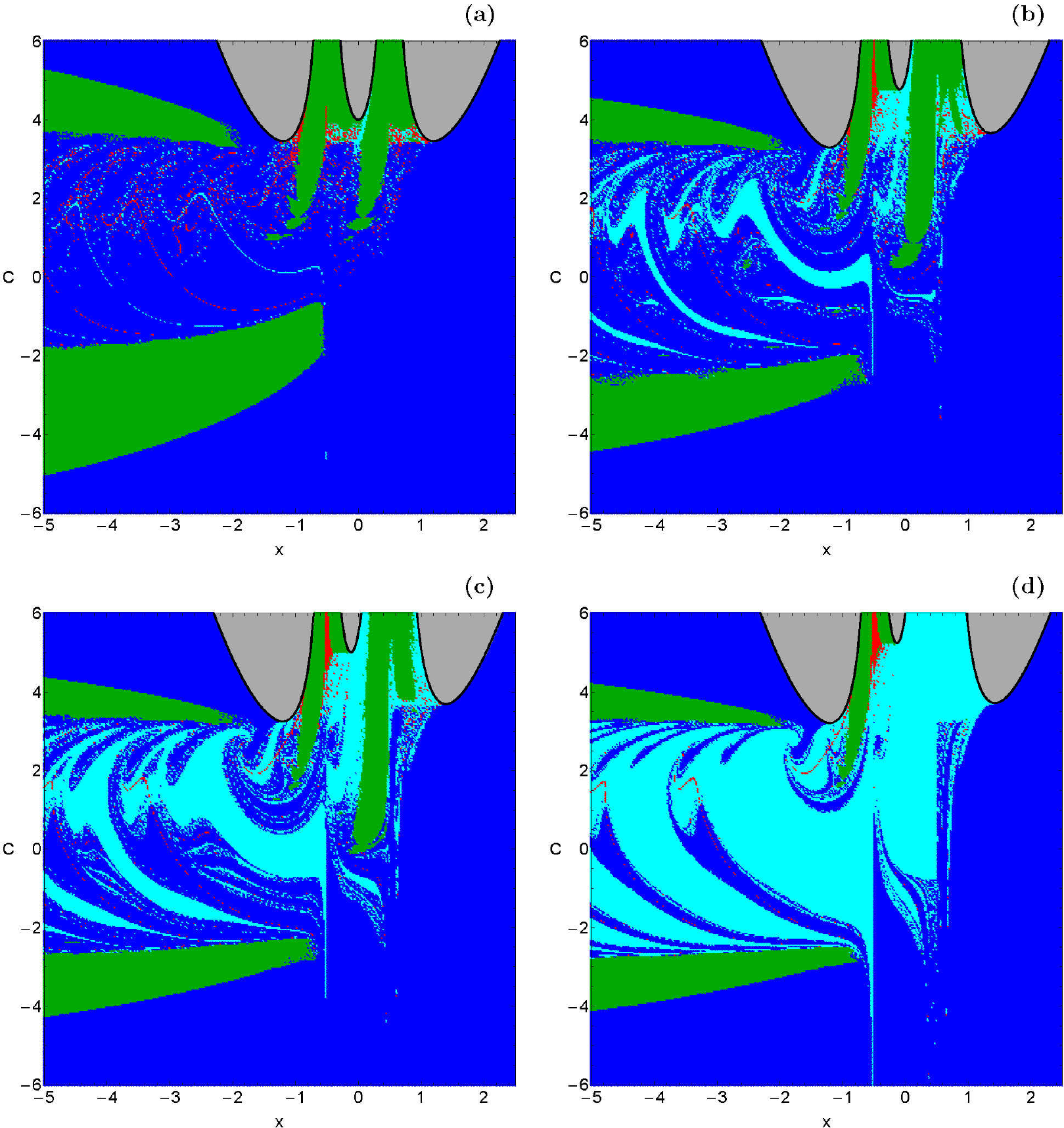}}
\caption{Orbital structure of the $(x,C)$ plane when (a-upper left): $p = 1$; (b-upper right): $p = 1.6$; (c-lower left): $p = 1.8$; (d-lower right): $p = 1.95$. The color code is the same as in Fig. \ref{hr1}.}
\label{xC}
\end{figure*}

\begin{figure*}[!t]
\centering
\resizebox{\hsize}{!}{\includegraphics{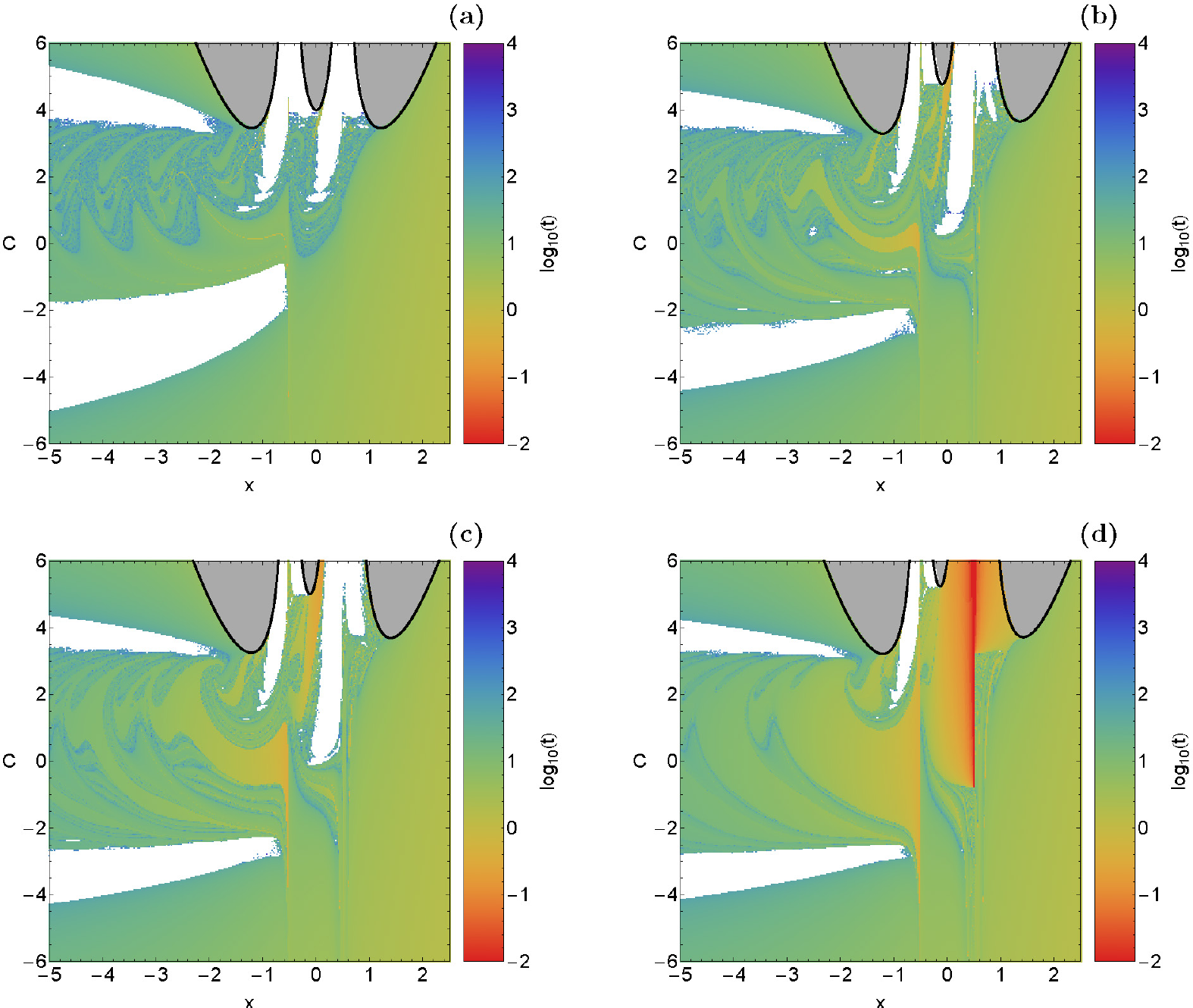}}
\caption{Distribution of the escape and collision time of the orbits on the $(x,C)$ plane for the values of $p$ of Fig. \ref{xC}.}
\label{xCt}
\end{figure*}

\begin{figure*}[!t]
\centering
\resizebox{\hsize}{!}{\includegraphics{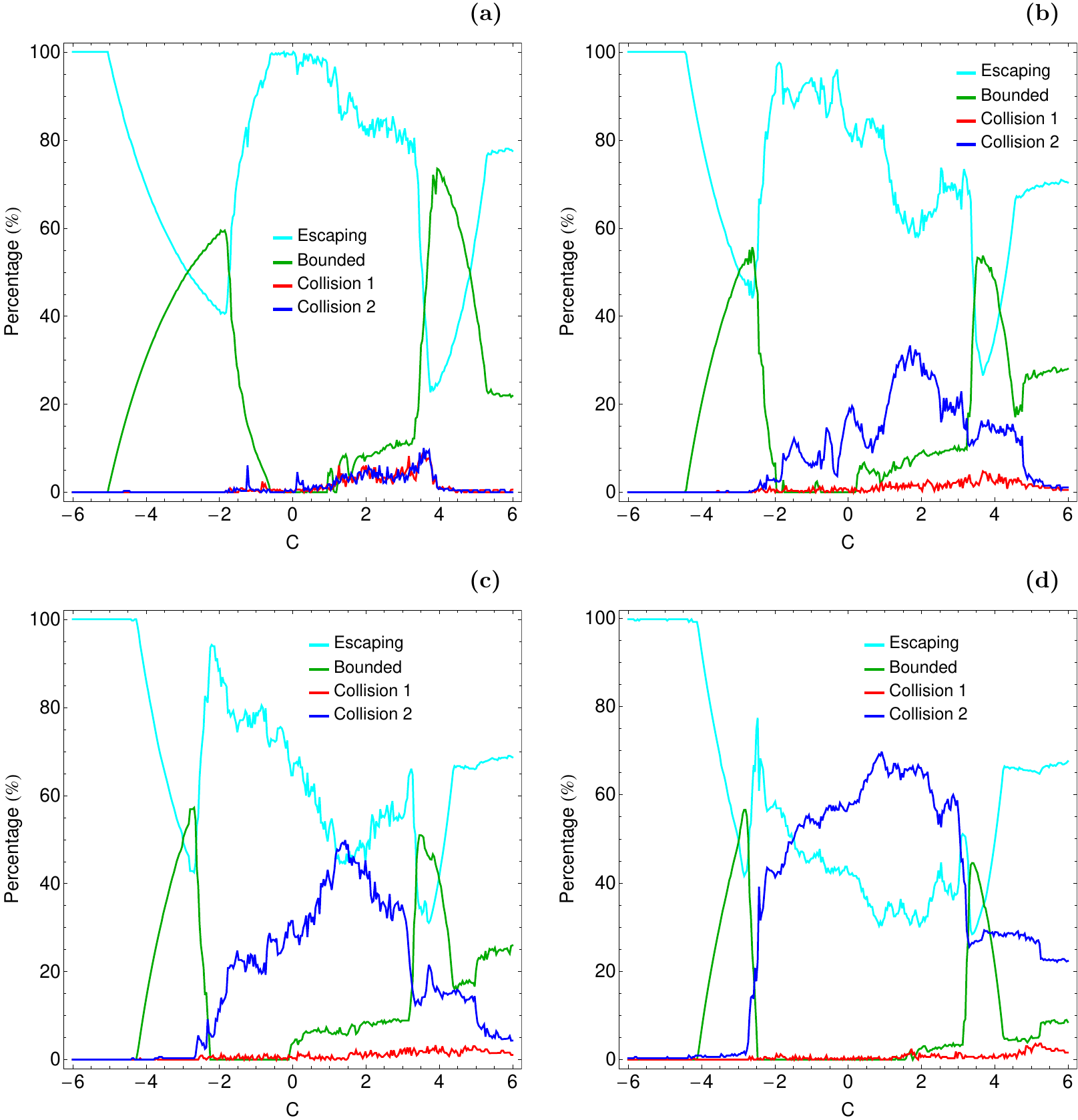}}
\caption{Evolution of the percentages of escaping, regular and collision orbits on the $(x,C)$-plane as a function of the value of the Jacobi constant $C$. (a-upper left): $p = 1$; (b-upper right): $p = 1.6$; (c-lower left): $p = 1.8$; (d-lower right): $p = 1.95$.}
\label{pC}
\end{figure*}

\begin{figure*}[!t]
\centering
\resizebox{\hsize}{!}{\includegraphics{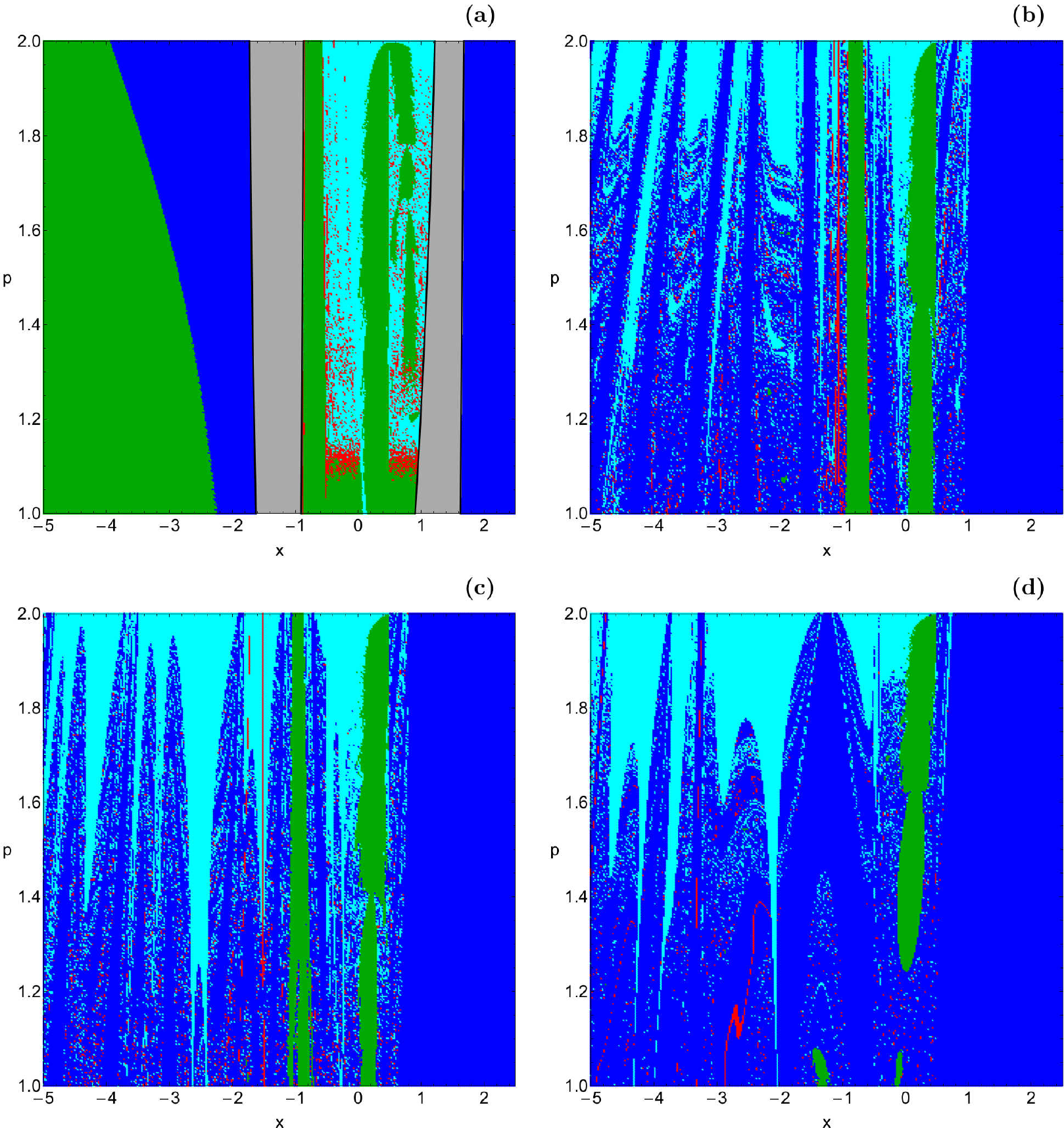}}
\caption{Orbital structure of the $(x,p)$ plane when (a-upper left): $C = 4$; (b-upper right): $C = 3$; (c-lower left): $C = 2$; (d-lower right): $C = 1$. The color code is the same as in Fig. \ref{hr1}.}
\label{xp}
\end{figure*}

\begin{figure*}[!t]
\centering
\resizebox{\hsize}{!}{\includegraphics{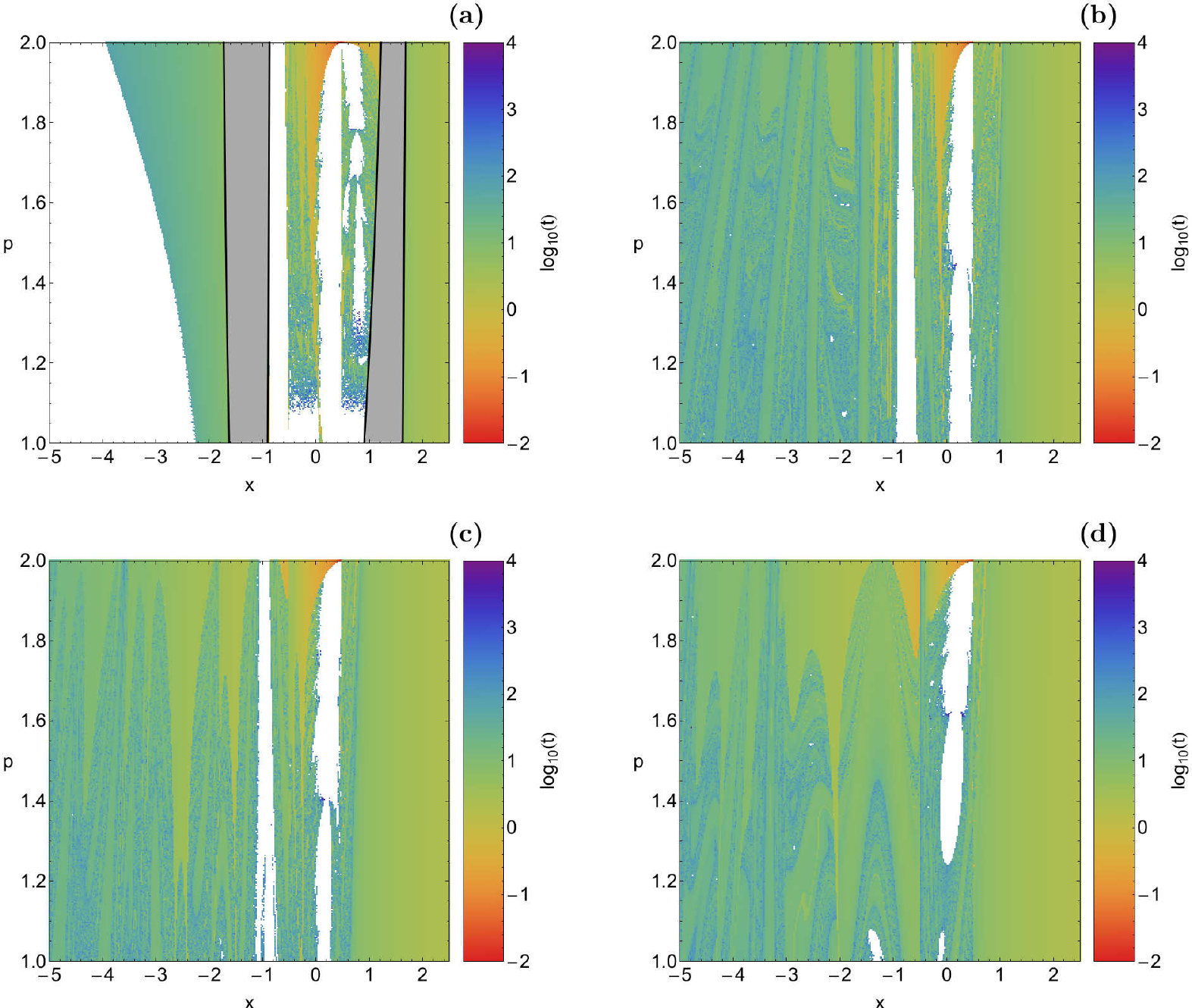}}
\caption{Distribution of the escape and collision time of the orbits on the $(x,p)$ plane for the values of the Jacobi constant $C$ of Fig. \ref{xp}.}
\label{xpt}
\end{figure*}

\begin{figure*}[!t]
\centering
\resizebox{\hsize}{!}{\includegraphics{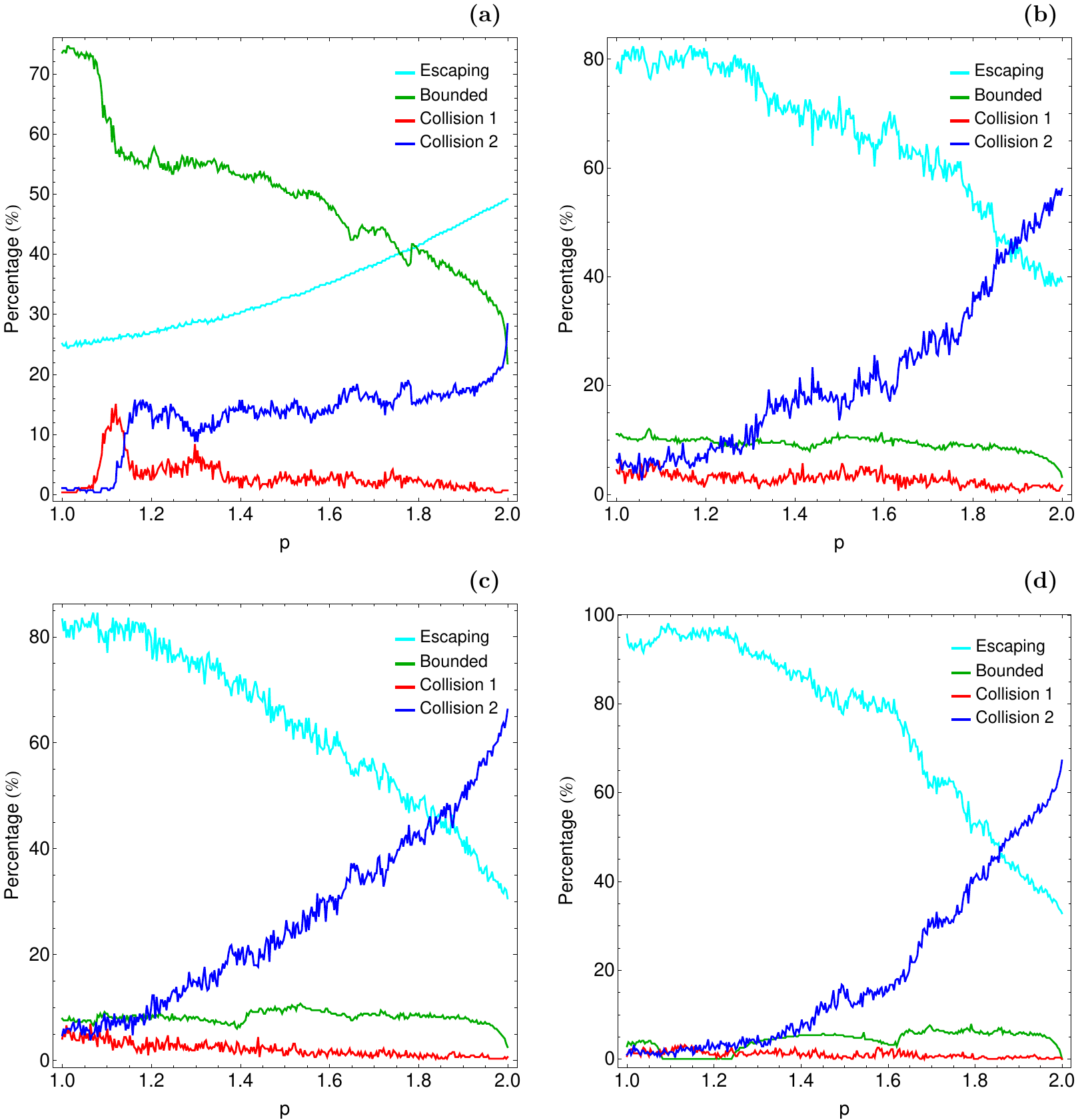}}
\caption{Evolution of the percentages of escaping, regular and collision orbits on the $(x,p)$-plane as a function of the power $p$ of the gravitational potential of primary 2. (a-upper left): $C = 4$; (b-upper right): $C = 3$; (c-lower left): $C = 2$; (d-lower right): $C = 1$.}
\label{pp}
\end{figure*}

\subsection{An overview analysis}
\label{over}

The color-coded OTDs in the configuration $(x,y)$ space, presented in the previous subsections, provide sufficient information on the phase space mixing however, for only a fixed value of the Jacobi constant (or the total orbital energy) and also for orbits that traverse the surface of section retrogradely. H\'{e}non \cite{H69}, introduced a new type of plane which can provide information not only about stability and chaotic regions but also about areas of bounded and unbounded motion using the section $y = \dot{x} = 0$, $\dot{y} > 0$ (see also \cite{BBS08}). In other words, all the initial conditions of the orbits of the test particles are launched from the $x$-axis with $x = x_0$, parallel to the $y$-axis $(y = 0)$. Consequently, in contrast to the previously discussed types of planes, only orbits with pericenters on the $x$-axis are included and therefore, the value of the Jacobi constant $C$ can now be used as an ordinate. In this way, we can monitor how the energy influences the overall orbital structure of our dynamical system using a continuous spectrum of Jacobi constants rather than few discrete values. In Fig. \ref{xC}(a-d) we present the orbital structure of the $(x,C)$ plane for four values of the power $p$ of the gravitational potential when $C \in [-6,6]$, while in Fig. \ref{xCt}(a-d) the distribution of the corresponding escape and collision times of the orbits is depicted. The black solid line in Fig. \ref{xC}(a-d) is the limiting curve which distinguishes between regions of allowed and forbidden motion and is defined as
\begin{equation}
f_1(x,C) = 2\Omega(x,y = 0) = C.
\label{zvc}
\end{equation}

We can observe the presence of several types of bounded orbits around the two primary bodies. Being more precise, on both sides of the primaries we identify stability islands corresponding to both direct (counterclockwise) and retrograde (clockwise) quasi-periodic orbits. It is seen that a large portion of the exterior region, that is for $x < x(L_3)$ and $x > x(L_2)$, a large portion of the $(x,C)$ plane is covered by initial conditions of escaping orbits however, at the left-hand side of the same plane two stability islands of regular orbits that circulate around both primaries are observed. We also see that collision basins leak outside the interior region, mainly outside $L_3$, and create complicated spiral shapes in the exterior region. It should be pointed out that in the blow-ups of the diagrams several additional very small islands of stability have been identified\footnote{An infinite number of regions of (stable) quasi-periodic (or small scale chaotic) motion is expected from classical chaos theory.}.

As the value of the power $p$ of the gravitational potential increases (which means that the gravity of primary 2 becomes stronger) the structure of the $(x,C)$ planes exhibits the following changes: (i) The collision basins to primary 1 located in the exterior region gradually weakens; (ii) The area of the stability islands around primary 1 remains almost unperturbed. According to Broucke's classification \cite{B68} the periodic orbits around the primaries belong to the families $C$ (at the left side of the primary) and $H_1$ (at the right side of the primary), while \cite{SS00} proved for the planar Hill's problem that the stability regions of the $C$ family are  more stable than those of the $H_1$ family; (iii) The stability islands around primary body 2 disappear only when $p = 1.95$. The phenomenon that stability islands can appear and disappear as a dynamical parameter is changed has also been reported in earlier paper (e.g., \cite{BBS06,dAT14}); (iv) The collision basins to primary 2 increase as the value of the power $p$ increases; (vi) The area of the two stability islands in the exterior region is reduced; (vii) Another interesting phenomenon is the fact that as the gravity of primary 2 becomes stronger the fractality of the $(x,C)$ plane reduces and the boundaries between escaping and collision basins appear to become smoother. It should be emphasized that the fractality of the structures was not measured by computing the corresponding fractal dimension.

An interesting aspect would be to monitor the evolution of the percentages of the different types of orbits as a function of the Jacobi constant $C$ for the $(x,C)$ planes shown in Figs. \ref{xC}(a-d). Our results are presented in Figs. \ref{pC}(a-d). Looking at four cases we see that escaping orbits dominate at very low $(C < -2)$ and at very high $(C > 4.5)$ values of the Jacobi constant, regardless the particular value of the power $p$ of the gravitational potential of primary 2. For intermediate values of the Jacobi constant the rate of escaping orbits fluctuate, however its value reduces with increasing $p$. The amount of collision orbits to primary 1 is always very small, less than 5\%, and it seems that it is not influenced by the increase of the gravitational field. On the other hand the percentage of collision orbits to primary 2 is highly affected by the increase of $p$. In particular it is seen that the rate of collision orbits to primary 2 gradually increases with increasing $p$, especially in the energy interval $-2.5 < C < 3.5$. When $p = 2$ this type of orbits is the most populated type of orbits in the energy range $-2.5 < C < 3.5$ since the corresponding rate is more than 50\%. At the same energy range the percentage of bounded motion is very low, while two peaks of the same type of motion are present at about $C = -3$ and $C = 4$. Taking into consideration all the above-mentioned analysis we may say that in the $(x,C)$ planes the power $p$ of the gravitational potential affects mostly escaping and collision orbits to primary 2, while the influence to the rest type of orbits is weaker.

It would be very illuminating if we had a more complete view of how the power $p$ of the gravitational potential influences the nature of orbits. In order to obtain this we follow a similar numerical approach to that explained before for the Jacobi constant thus examining now a continuous spectrum of $p$ values. In particular, we use again the section $y = \dot{x} = 0$, $\dot{y} > 0$, launching orbits once more from the $x$-axis with $x = x_0$, parallel to the $y$-axis, for a specific value of the Jacobi constant $C$. This allow us to construct again a two-dimensional plane in which the $x$ coordinate of orbits is the abscissa, while the power $p$ is the ordinate. In Fig. \ref{xp}(a-d) we present the orbital structure of the $(x,p)$-plane when $p \in [1,2)$ for four values of the Jacobi constant, while in Fig. \ref{xpt}(a-d) the distribution of the corresponding escape and collision times of orbits is depicted. The outermost black solid line is the limiting curve which this time is given by
\begin{equation}
f_2(x;C,p) = 2\Omega(x,y = 0) = C.
\label{zvc2}
\end{equation}

Looking at Fig. \ref{xp}(a-d) we see that as the value of the power $p$ increases the following phenomena take place: (i) for low values of the Jacobi constant $(C < 4)$ the amount of escaping orbits decreases apart from the case where $C = 4$ at which the amount of escaping orbits increases; (ii) the collision basins to primary 2 significantly increase; (iii) the initial conditions of orbits which collide with primary 1 seem to decrease; (iv) the extent of bounded basins is reduced and when $p = 1.95$ the stability islands corresponding to motion around primary 2 completely disappear; (vi) the fractality of the $(x,p)$ planes decreases and the basins boundaries are more smooth.

We close our numerical investigation by presenting the evolution of the percentages of all types of orbits as a function of the power $p$. Fig. \ref{pp}(a-d) contains the diagrams corresponding to the $(x,p)$ planes of Fig. \ref{xp}(a-d), respectively. We observe that evolution of the percentages in panels (b)-(d) is very similar. In particular, for low values of $p$ escaping orbits is the most populated type of orbits occupying more than 70\% of the $(x,p)$ planes. As the value of $p$ increases however, the rate of escaping orbits is constantly reduced and it reaches at about 40\% (or lower) when $p$ approaches 2. The evolution of the percentage of collision orbits to primary 2 is completely different, as their rate is heavily increasing. At the highest value of $p$ collision orbits to primary 2 cover more than 60\% of the $(x,p)$ planes for $C < 4$. The rates of collision orbits to primary 1 and that of bounded orbits fluctuate throughout at relatively low values below 10\%. The case where $C = 4$ which is presented in panel (a) is quite different. Here bounded orbits dominate in the interval $1 \leq p < 1.8$, while escaping orbits take over the $(x,p)$ plane for higher values of $p$. The percentages of collision orbits to primaries 1 and 2 seem to fluctuate at about 2\% and 15\%, respectively. The results shown in Fig. \ref{pp}(a-d) suggest that in the $(x,p)$ planes the power $p$ of the gravitational potential affects, more or less, all types of orbits except for the collision orbits to primary 1 with the classical Newtonian gravity.

\section{Discussion and conclusions}
\label{disc}

The main scope of this numerical investigation was to unveil how the power of the gravitational potential influences the character of orbits in the classical planar circular restricted three-body problem (PCRTBP) with strong gravitational field. After conducting an extensive and thorough numerical investigation we managed to distinguish between bounded, escaping and collision orbits and we also located the basins of escape and collision, finding also correlations with the corresponding escape and collision times. Our numerical results strongly suggest that the power of the potential plays a very important role in the nature of the test's particle motion under the gravitational field of the two primaries. To our knowledge, this is the first detailed and systematic numerical analysis on the influence of the gravitational field on the character of orbits in PCRTBP and this is exactly the novelty and the contribution of the current work.

For several values of the power of the potential in the last four Hill's regions configurations we defined dense uniform grids of $1024 \times 1024$ initial conditions regularly distributed on the $\dot{\phi} < 0$ part of the configuration $(x,y)$ plane inside the area allowed by the value of the Jacobi constant (or in other words by the value of the total orbital energy). All orbits were launched with initial conditions inside the scattering region, which in our case was a square grid with $-2\leq x,y \leq 2$. For the numerical integration of the orbits in each type of grid, we needed about between 10 hours and 2 days of CPU time on a Quad-Core i7 2.4 GHz PC, depending on the escape and collision rates of orbits in each case. For each initial condition, the maximum time of the numerical integration was set to be equal to $10^4$ time units however, when a test particle escaped or collided with one of the two primaries the numerical integration was effectively ended and proceeded to the next available initial condition.

In this study we provide quantitative information regarding the escape and collision dynamics in the PCRTBP with strong gravitational field. The main outcomes of our numerical research can be summarized as follows:
\begin{enumerate}
 \item In the second Hill's regions configurations, where the transport channel near $L_1$ is open, we found that the increase of the power of the gravitational potential of primary 2 affects mostly the region around the same primary.
 \item In the third Hill's regions configurations, where the escape channel near $L_2$ is open, the power $p$ influences all the interior region since it was observed that the collision basins to primary 1 are reduced with increasing $p$.
 \item In the fourth and fifth Hill's regions configurations we seen that the influence of $p$ expands to the entire configuration $(x,y)$ space.
 \item In all types of planes two important phenomena take place with increasing $p$: (i) the extent of the collision basins to primary 2 heavily increases, while the stability regions corresponding to bounded motion around primary 2 disappear at relatively high values of $p$.
 \item Our calculations suggest that in most cases the orbital structure of the regions around primary 1 remains almost unperturbed by the shifting on the power $p$ of the gravitational potential.
 \item The degree of fractality on the several types of planes seems to be reduced with increasing $p$ as the boundaries between the different types of basins become smoother.
\end{enumerate}

Judging by the detailed and novel outcomes we may say that our task has been successfully completed. We hope that the present numerical analysis and the corresponding results to be useful in the field of escape and collision dynamics in the PCRTBP with strong gravitational field. The results as well as the conclusions of the present research are considered, as an initial effort and also as a promising step in the task of understanding the orbital structure in this interesting problem of dynamics. Taking into account that our outcomes are encouraging, it is in our future plans to properly modify our dynamical model in order to expand our investigation into three dimensions and explore the entire six-dimensional phase thus revealing the influence of the power of the gravitational potential on the character of orbits. In addition, it would be very interesting to obtain the network of periodic orbits (stable and unstable) and determine how they are influenced by the Jacobi constant $C$ and the power $p$ of the gravitational potential.

\section*{Acknowledgments}

I would like to express my warmest thanks to the two referees for the careful reading of the manuscript and for all the apt suggestions and comments which allowed us to improve both the quality and the clarity of the paper.

\section*{Appendix: Existence and stability of circular orbits in the generalized Kepler problem}
\label{apx}

It is well known that in the classical Newtonian gravitational potential $V = - k/r^n$, where $n = 1$, a circular orbit always exists. When $n \neq 1$ we have the case of the generalized Kepler problem. According to theoretical mechanics a necessary condition for the existence of circular orbits is the central potential to be attractive. In the following we shall demonstrate for which values of $n$ we stable or unstable circular solutions.

\begin{figure}[H]
\begin{center}
\includegraphics[width=\hsize]{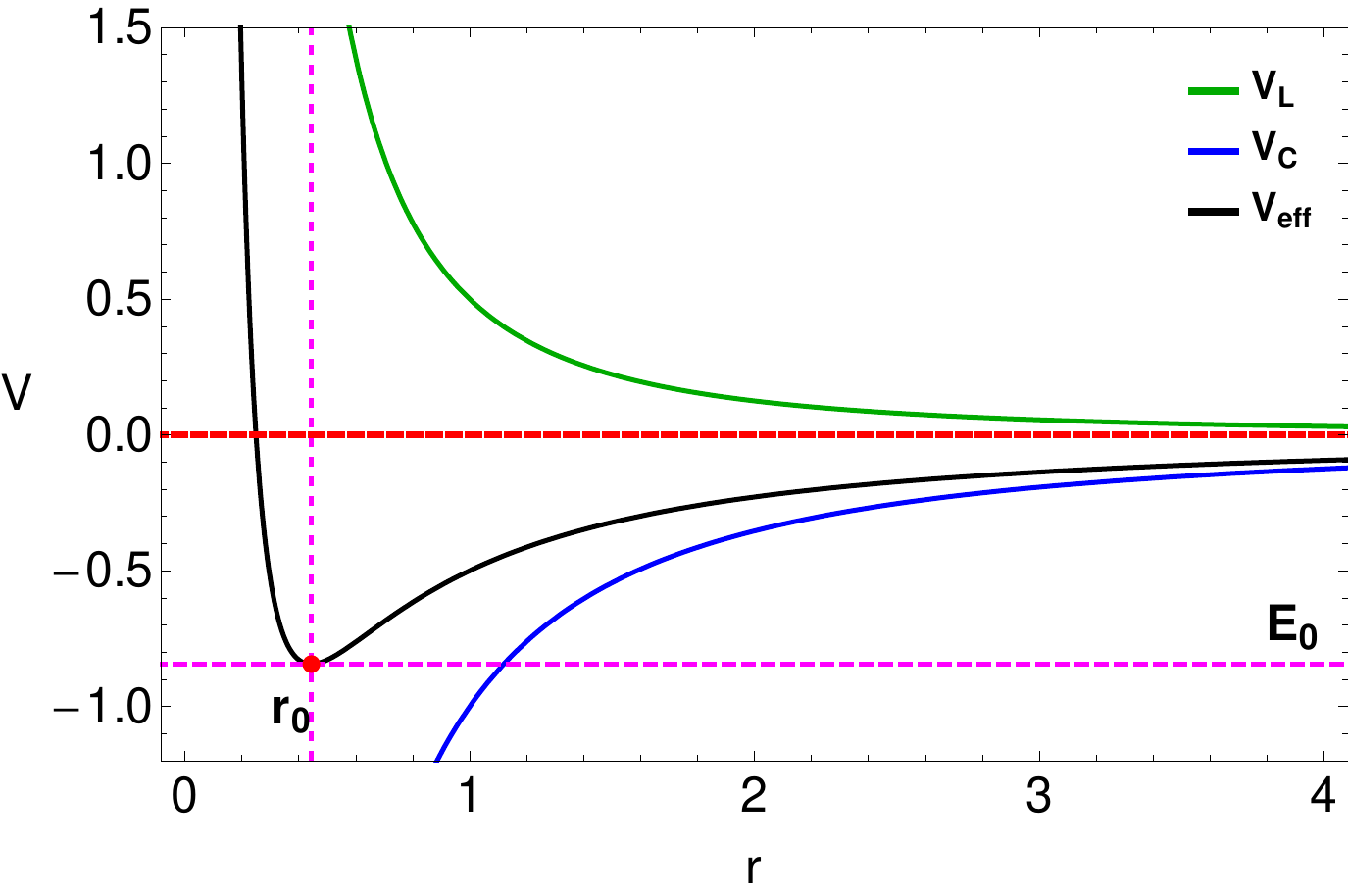}
\end{center}
\caption{Evolution of $V_{\rm eff}$ (black), $V_C$ (blue), and $V_L$ (green) as a function of radius $r$, when $k = m = L = 1$, and $n = 3/2$. The vertical and the horizontal magenta dashed lines indicate the radius $r_0$ and the corresponding energy $E_0$ of the circular orbit, respectively.}
\label{pots}
\end{figure}

The total effective potential $V_{\rm eff}$ in a central force field is the sum of the central potential $V_C$ and the centrifugal term $V_L$.
\begin{equation}
V_{\rm eff}(r) = V_C(r) + V_L(r) = - \frac{k}{r^n} + \frac{L^2}{2mr^2},
\label{veff}
\end{equation}
where $n > 0$, $k > 0$, and $m > 0$. In Fig. \ref{pots} we present a plot of $V_{\rm eff}$, $V_C$, and $V_L$ when $k = m = L = 1$ and $n =3/2$.

The radius $r_0$ of the circular orbit, if exists, is obtained by setting the first derivative of the total effective potential equal to zero. Therefore we have
\begin{equation}
r_0(n) = \left(\frac{mkn}{L^2}\right)^{\frac{1}{n-2}} > 0.
\label{r0}
\end{equation}
For $n = 3/2$ we derive that $r_0 = \frac{4L^2}{9m^2k^2}$. Looking at Eq. (\ref{r0}) it becomes evident that circular solutions are indeed permissible for every value of $n$ except for $n = 2$.

The energy of the circular orbit, $E_0$, can easily be found if we insert the expression (\ref{r0}) of $r_0$ into Eq. (\ref{veff}). After applying elementary calculations we obtain
\begin{equation}
E_0(n) = - k \left(\frac{mkn}{L^2}\right)^{\frac{n}{2-n}}\left(1 - \frac{n}{2}\right).
\label{E0}
\end{equation}
The last term of Eq. (\ref{E0}), that is $\left(1 - \frac{n}{2}\right)$, suggests that when $0 < n <2$ $E_0$ is negative, while for $n > 2$ $E_0$ is positive. In other words, for $0 < n <2$ the equilibrium point of $V_{\rm eff}$ at $r_0$ is a global minimum, while for $n > 2$ it is a global maximum. Here it should be noticed that the sign of the energy at the equilibrium point already implies the stability of the circular orbits.

Our next task is to determine for which values of $n$ the corresponding circular orbit is stable and for which becomes unstable. It is very easy to prove that in a central force field the necessary and also sufficient condition for a circular orbit with radius $r_0$ to be stable is the following
\begin{equation}
S(r_0) = \frac{F'(r_0)}{F(r_0)} + \frac{3}{r_0} > 0,
\label{stab}
\end{equation}
where of course
\begin{equation}
F(r) = - \frac{dV_C(r)}{dr} = - n \frac{k}{r^{n+1}},
\label{for}
\end{equation}
and
\begin{equation}
F'(r) = \frac{dF(r)}{dr} = n\left(n + 1\right) \frac{k}{r^{n+2}}.
\label{dfor}
\end{equation}

\begin{figure}[H]
\begin{center}
\includegraphics[width=\hsize]{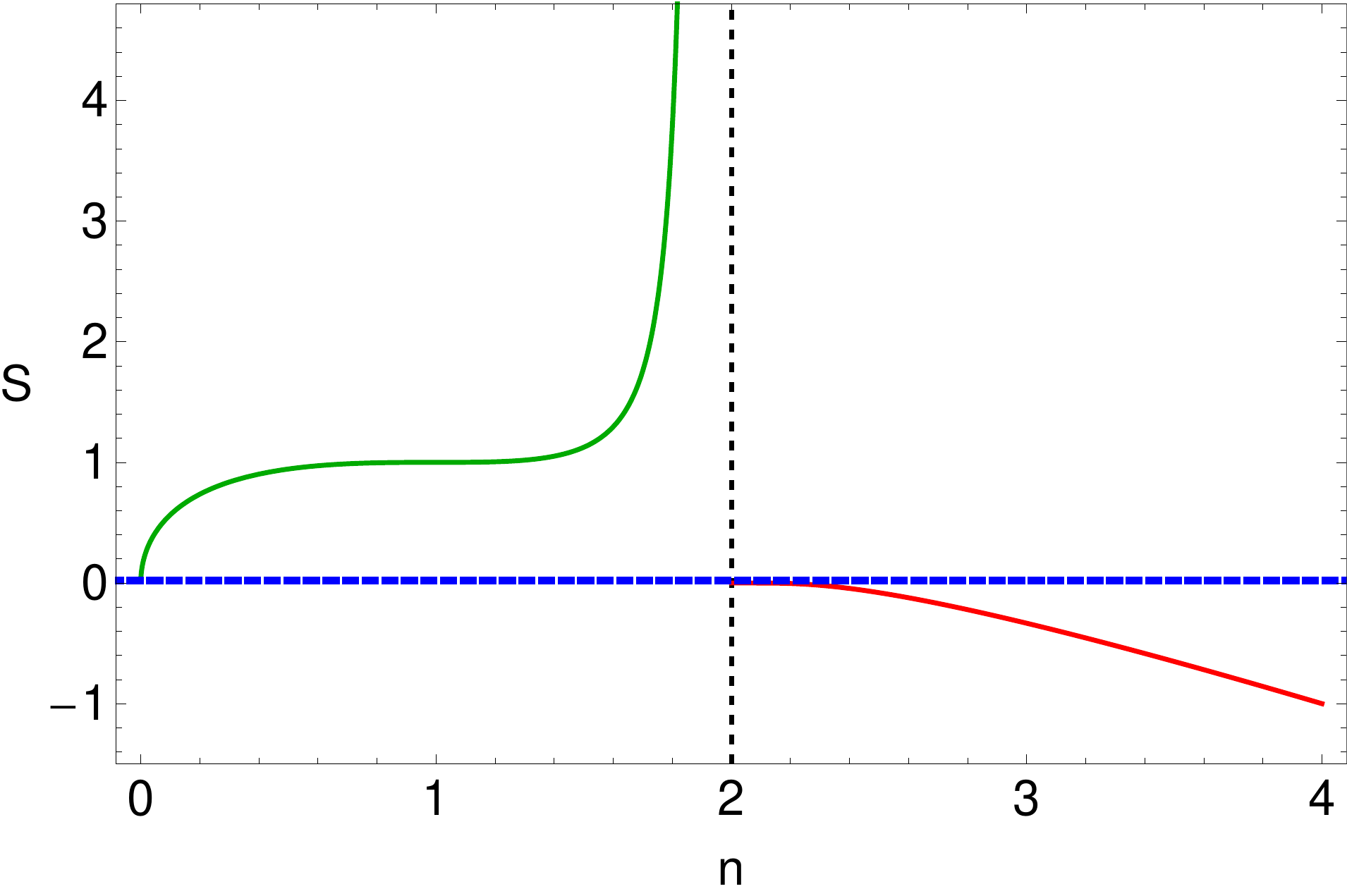}
\end{center}
\caption{Evolution of $S$ as a function of the power $n$ of the gravitational central potential $V_C$, when $k = m = L = 1$. The horizontal blue dashed line denotes the threshold $S = 0$, which distinguishes between stable and unstable circular orbits. Stable circular orbits exist only in the interval $n \in (0,2)$, while for $n > 2$ circular orbits are permissible however they are all unstable.}
\label{si}
\end{figure}

Inserting Eqs. (\ref{r0}), (\ref{for}), and (\ref{dfor}) into (\ref{stab}) we obtain $S$ as a function of $n$ as follows
\begin{equation}
S(n) = - \left(n - 2\right) \left(\frac{mkn}{L^2}\right)^{\frac{1}{2-n}}.
\label{stab2}
\end{equation}
In Fig. \ref{si} we illustrate the evolution of $S(n)$, when $k = m = L = 1$. It is seen, that in the interval $n \in (0,2)$ $S > 0$ and therefore stable circular orbits exist. On the other hand, for $n > 2$ circular orbits do exist however they are unstable since $S < 0$. In this paper we investigate the nature of orbits in the planar circular restricted three-body problem where one of the primaries has a strong gravitational field. We consider only cases where $n$ varies in the interval $(1,2)$, where according to Eq. (\ref{stab2}) stable Keplerian orbits are indeed permissible.

\end{document}